\documentclass[10pt,journal,compsoc]{IEEEtran}
\usepackage{booktabs} 
\usepackage[table]{xcolor}
\usepackage{xspace}
\usepackage{url}
\usepackage{balance} 
\usepackage{tabulary}
\usepackage{array}
\usepackage{graphicx}
\usepackage{hyperref}
\hypersetup{hidelinks=true} 
\usepackage[ruled,lined,linesnumbered,vlined,algo2e]{algorithm2e}
\usepackage{lipsum}
\usepackage{subfigure}
\usepackage[T1]{fontenc}
\usepackage{amsmath, amsthm, amssymb}
\usepackage[ansinew]{inputenc}
\usepackage[many]{tcolorbox}
\usepackage{color}
\usepackage{xcolor, colortbl}
\definecolor{grayone}{gray}{.9}
\definecolor{graytwo}{gray}{.7}
\usepackage{extarrows}
\usepackage{multirow}
\usepackage{adjustbox}
\usepackage{tabularx}

\newcommand{\sen}[1]{\textcolor{black}{#1}}

\newcommand{\revised}[1]{\textcolor{black}{#1}}

\newcommand{\changed}[1]{\textcolor{black}{#1}}

\newcommand{\ours}{Xbot\xspace}
\newcommand{\their}{AT\_{Monkey}\xspace}

\newcommand{\rOne}[1]{\textcolor{black}{#1}}

\makeatletter
\newcommand*\bigcdot{\mathpalette\bigcdot@{1}}
\newcommand*\bigcdot@[2]{\mathbin{\vcenter{\hbox{\scalebox{#2}{$\m@th#1\bullet$}}}}}
\makeatother

\begin{document}

\title{Accessible or Not? An Empirical Investigation of Android App Accessibility}

\author{
        Sen Chen,
        Chunyang Chen, 
        Lingling Fan,
        Mingming Fan, 
        Xian Zhan,
        and~Yang Liu
\IEEEcompsocitemizethanks{
	
	\IEEEcompsocthanksitem
	Sen Chen is with College of Intelligence and Computing, Tianjin University, China. 
	Email: senchen@tju.edu.cn.
	Chunyang Chen is with Monash University, Australia. 
	Email: chunyang.chen@monash.edu.
	Lingling Fan is with College of Cyber Science, Nankai University, China.
	Email: linglingfan@nankai.edu.cn.
	Mingming Fan is with The Hong Kong University of Science and Technology. Email: mingmingfan@ust.hk. 
	Xian Zhan is with The Hong Kong Polytechnic University. Email: chichoxian@gmail.com.
	Yang Liu is with School of Computer Science and Engineering, Nanyang Technological University.
	Email: yangliu@ntu.edu.sg.
	\IEEEcompsocthanksitem Chunyang Chen and Lingling Fan are the corresponding authors.
}
}

\markboth{Journal of \LaTeX\ Class Files,~Vol.~xx, No.~xx, xx~2021}%
{Chen \MakeLowercase{\textit{et al.}}: Accessible or Not? An Empirical Investigation of Android App Accessibility}

\IEEEtitleabstractindextext{%

\begin{abstract}
Mobile apps provide new opportunities to people with disabilities to act independently in the world. Following the law of the US, EU, mobile OS vendors such as Google and Apple have included accessibility features in their mobile systems and provide a set of guidelines and toolsets for ensuring mobile app accessibility. Motivated by this trend, researchers have conducted empirical studies by using the inaccessibility issue rate of each page (i.e., screen level) to represent the characteristics of mobile app accessibility. However, there still lacks an empirical investigation directly focusing on the issues themselves (i.e., issue level) to unveil more fine-grained findings, due to the lack of an effective issue detection method and a relatively comprehensive dataset of issues.

To fill in this literature gap, we first propose an automated app page exploration tool, named \ours, to facilitate app accessibility testing and automatically collect accessibility issues by leveraging the instrumentation technique and static program analysis. Owing to the relatively high activity coverage (around 80\%) achieved by \ours when exploring apps, Xbot achieves better performance on accessibility issue collection than existing testing tools such as Google Monkey. With \ours, we are able to collect a relatively comprehensive accessibility issue dataset and finally collect 86,767 issues from 2,270 unique apps including both closed-source and open-source apps, based on which we further carry out an empirical study from the perspective of accessibility issues themselves to investigate novel characteristics of accessibility issues. Specifically, we extensively investigate these issues by checking 1) the overall severity of issues with multiple criteria, 2) the in-depth {relation} between issue types and app categories, GUI component types, 3) the frequent issue patterns quantitatively, and 4) the fixing status of accessibility issues. Finally, we highlight some insights to the community and hope to raise the attention to maintaining mobile app accessibility for users especially the elderly and disabled.
\end{abstract}

}

\maketitle

\IEEEdisplaynontitleabstractindextext

\IEEEpeerreviewmaketitle

\IEEEraisesectionheading{\section{Introduction}\label{sec:introduction}}

\begin{figure}
\centering
\includegraphics[width=0.5\textwidth]{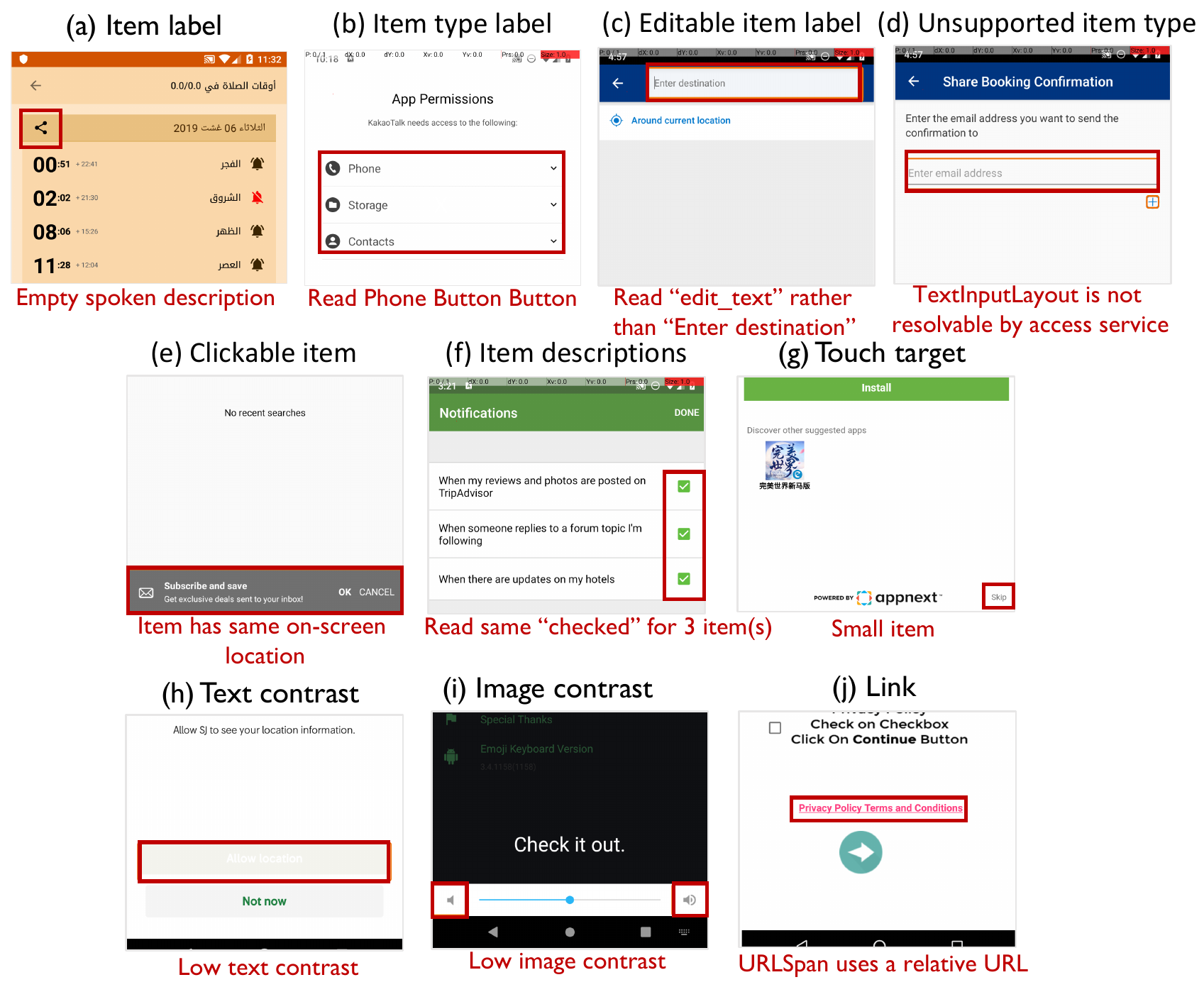}
\caption{Examples of accessibility issues with brief descriptions}
\label{fig:items}
\end{figure}

\IEEEPARstart{A}{s} mobile applications (apps) are increasingly embedded into people's daily lives, ensuring their accessibility to a broader range of users has gained increasing attention from both industry and governments. For example, leading IT companies (e.g, Apple, Google, IBM, and Microsoft) have established their accessibility teams~\cite{AppleAccessibility, MicrosoftAccessibility, IBMAccessibility, FacebookAccessibility} and governments have established laws to help eliminate barriers in electronic and information technology for people with disabilities~\cite{web:EuropeLaw,web:USlaw}.
Although there are many accessibility guidelines for mobile app development (e.g., \cite{WCAG,BBC}), it is challenging for mobile apps designers and developers who often neither have disabilities themselves nor have training in user experience (UX) and accessibility, to figure out how to discover potential \textit{accessibility issues}\footnote{Accessibility issue refers to issues that make apps less accessible to people with disabilities such as blind users when they are using mobile phones. Fig.~\ref{fig:items} shows some examples of accessibility issues.} for a wide range of disabilities, and apply accessibility guidelines to effectively address the issues~\cite{Trewin:2010:ACT:1805986.1806029,Bigham:2010:ADE:1878803.1878812}. 
Furthermore, in practice, many small start-up companies often have limited, if any, professional user interface (UI)/UX designers with expertise to address accessibility related issues \cite{hokkanen2015ux}. For example, Fig.~\ref{fig:items} shows some accessibility issues that frequently occur in mobile apps, which cause problems to the elderly and disabled (e.g., item label missing~\cite{chen2020unblind,ross2020epidemiology} causing spoken errors when using TalkBack~\cite{talkback} for blind users in Fig.~\ref{fig:items}(a)), some issues are even inaccessible to users without disabilities, e.g., low text contrast in Fig.~\ref{fig:items}(h) (details in \S~\ref{subsec:issues}). 
	
To improve app accessibility, some researchers from the academia and industry both paid more attention to understanding the status of app accessibility and mining the characteristics of introduced issues~\cite{serra2015accessibility,ross2018examining, yan2019current, vendome2019can, alshaybanaccessibility} to reduce accessibility issues. However, the existing static rule-based checking methods (e.g., Lint~\cite{lint}, Espresso~\cite{web:androidEspresso}, Robolectric~\cite{web:androidRobolectric}) have been demonstrated to be ineffective and time-consuming for detecting mobile accessibility issues~\cite{eler2018automated,fan2018efficiently,alshaybanaccessibility,fan2018large,yan2019current,ross2018examining}. On the other hand, some big companies such as Google provide some accessibility testing tools (e.g., Google Accessibility Scanner~\cite{scanner} and IBM AbilityLab Mobile Accessibility Checker~\cite{IBMAccessibility}) for detecting accessibility issues on each UI page of apps, which requires human intervention. To make app accessibility testing tools fully automated, researchers~\cite{yan2019current, alshaybanaccessibility} adopt dynamic app testing tools (e.g., Google Monkey~\cite{monkey}) to dynamically explore the app and feed the explored UI pages to the accessibility testing tools for detecting accessibility issues. Based on the collected accessibility issues, they carry out empirical studies in terms of the prevalence of accessibility issues. However, the latest related work~\cite{alshaybanaccessibility} acknowledged that existing testing tool (i.e., Google Monkey) can only achieve a low activity coverage (around 40\%), and they can only obtain a limited number of issues for each app. Their analysis is based on a limited dataset for each app, which is enough for the study at the \textbf{screen level} (i.e., using and measuring the inaccessibility issue rate of each screen), but hard to reveal more fine-grained findings at the \textbf{issue level} (i.e., directly focusing on the issues themselves). Therefore, to empirically investigate accessibility issues directly, first of all, it is necessary to simulate user interactions to explore as many app pages as possible and further collect a large-scale and relatively comprehensive dataset of app accessibility issues. With such a dataset, we aim to conduct an empirical study to reveal more fine-grained insights from the perspective of issues themselves.

To achieve this goal, two challenges need to be overcome: (1) Firstly, there lacks an effective tool to automatically explore app UI pages with high activity coverage. High activity coverage can help simulate various user interactions. To conduct an empirical investigation of accessibility issues, it is essential to check as many activities as possible to collect accessibility issues. (2) Secondly, there lacks a large-scale and relatively comprehensive dataset about real-world app accessibility issues for the further empirical study and investigation. Enabling app accessibility analysis requires a comprehensive set of issues including the user interface screenshots, the detailed accessibility descriptions, the buggy front-end source code, and issue patches (if any).

To this end, we propose a novel tool named \textbf{\ours}, to automatically and effectively explore UI pages to facilitate accessibility testing and collect accessibility issues in apps. It leverages instrumentation and static program analysis techniques. \ours is demonstrated to achieve better performance than the existing data collection methods based on manual exploration and random testing exploration with Monkey in recent work~\cite{alshaybanaccessibility}. By leveraging \ours, we automatically assess 17,417 app pages from 2,270 apps and finally collect 86,767 accessibility issues, which is the largest dataset for app accessibility until now. We have released it along with the source code of Xbot on Github: \url{https://github.com/tjusenchen/Xbot}. We then carry out an empirical investigation of these accessibility issues from different dimensions by answering the following research questions:

\begin{itemize}
    \item \textbf{RQ1}: Can Xbot outperform the existing methods on app page exploration and issue collection when conducting accessibility testing?
    \item \textbf{RQ2}: What is the overall severity status of app accessibility at the issue level for both closed-source and open-source apps?
    \item \textbf{RQ3}: What are the {in-depth relations} between the accessibility issue types and app category, GUI component?
    \item \textbf{RQ4}: What are the quantitative characteristics of specific issues such as text or image contrast issues?
    \item \textbf{RQ5}: How many accessibility issues have been fixed during app version updates?
\end{itemize}
	
According to the investigation of app accessibility, we find that
(1) 89\% apps are overall suffering from severe accessibility problems for both open-source and closed-source apps, with 43 issues for each app and 6.5 issues for each page on average;
(2) most of the accessibility issues remain unfixed (96\%) according to the investigation on the multiple history versions, which is inconsistent with the previous study (47\% high fixing rate in the previous study vs. 4\% low fixing rate in our study), mainly due to the unsteady activity coverage of the underlying testing tools used by them.
(3) \textit{Touch target, Text contrast, Item label} are the top 3 issue types ranked by the number of issues. 5 types of GUI components (i.e., \textit{TextView, ImageView, Button, EditText, and ImageButton}) are often associated with accessibility issues; and (4) different issue types may have different frequency across different app categories such as the small size of touchable components in shopping apps, thus, app developers should take this feature into consideration to maintain their own apps' accessibility. More fine-grained findings can be found in Section~\ref{sec:new_study}.

In summary, we make the following contributions:
\begin{itemize}
\item A fully automated and effective app UI exploration tool\footnote{\url{https://github.com/tjusenchen/Xbot}} for dynamically scanning mobile app accessibility issues and collecting a relatively comprehensive dataset of issues for further studies. 

\item A comparative study to demonstrate the better performance on accessibility issue collection of our tool with others such as manual exploration and the existing dynamic methods by leveraging Google Monkey.

\item An in-depth and empirical study of accessibility issues based on our collected large-scale dataset, which unveils insights for the community to better understand the characteristics of issues and further improve mobile apps' accessibility.

\item A large-scale and reusable dataset~\cite{mysite} including {86,767} issues from 2,270 apps and their metadata (e.g., issue descriptions), which enables the community to further advance mobile app accessibility research. Meanwhile, the source code of Xbot is also released for the community.
\end{itemize}

\section{Preliminary}\label{sec:preliminary}
Apart from the 15\% population with disabilities who were born blind, or lost fine motor skills in an accident, most people may also have a short-term disability at some time that makes it difficult to use their mobile devices. For example, someone cannot use their hands because they are carrying a wiggly child, have experienced difficulties using the phone while wearing gloves when it is cold outside, or maybe have a hard time distinguishing items on the screen when it is bright outside. With so much of the population experiencing decreased vision, hearing, mobility, and cognitive function, developers should do their best to give everyone the best experience in their apps. The UN Convention on the Rights of Persons with Disabilities recognizes access to information and communications technologies, including the mobile apps, as a basic human right~\cite{web:UNaccessibility} and social justice~\cite{ladner2015design}.
	
In this section, we briefly introduce the definition of accessibility and the app accessibility issue types that detected by Google Accessibility Test Framework~\cite{framework} and Google Accessibility Scanner~\cite{scanner}. 

\subsection{Accessibility Guidelines}\label{subsec:guidelines}
W3C (World Wide Web Consortium), the main international standards organization for the World Wide Web has very clear web content accessibility guidelines (WCAG)~\cite{web:W3CwebAccessibility} for developing accessible websites which can be accessed by users with disabilities. Based on the web accessibility, they further develop the accessibility standards for mobile applications~\cite{web:W3CmobileAccessibility} by considering mobile characteristics such as touch screens, small screen size, usages in different settings like bright sunlight, etc. In addition to general accessibility guidelines, researchers have proposed accessibility guidelines for special populations, such as people with visual impairments \cite{Park:2014:TAM:2729485.2729491}, people with hearing impairments \cite{jaramillo2018approach}, people with Aphasia~\cite{Grellmann:2018:IMA:3234695.3241011}, or older adults \cite{diaz2014accessibility}.
	
At the same time, as the primary organizations that facilitate mobile technology and the app marketplace, Google and Apple also release their accessibility guidelines~\cite{google_document}, SDKs~\cite{framework}, and testing suites~\cite{suite} for mobile apps on Android and iOS platforms. Despite the importance of these guidelines, the guidelines are difficult for app designers or developers to comprehend and implement into app design~\cite{Clegg-Vinell:2014:IAR:2596695.2596717}. As a result, there is a need to facilitate the evaluation of accessibility issues of mobile apps using the guidelines. 

\subsection{App Accessibility Issues}\label{subsec:issues}
Following the accessibility guidelines provided by Google, we identify 10 kinds of accessibility issues.
We briefly describe each issue type and provide real examples in Fig.~\ref{fig:items} to illustrate what real accessibility issues are like in user interface pages. 
\begin{itemize}
    \item \textit{Item label} in Fig.~\ref{fig:items}(a) means views that a screen reader could focus and that have an empty spoken description.
    \item \textit{Item type label} in Fig.~\ref{fig:items}(b) means Views with a redundant description.
    \item \textit{Editable item label} in Fig.~\ref{fig:items}(c) means EditTexts and editable TextViews that have a non-empty contentDescription, thus a screen reader may read this attribute instead of the editable content when the user is navigating.
    \item \textit{Unsupported item type} in Fig.~\ref{fig:items}(d) means item types that are not supported by accessibility services.
    \item \textit{Clickable item} in Fig.~\ref{fig:items}(e) means more than one item share the same on-screen location.
    \item \textit{Item description} in Fig.~\ref{fig:items}(f) means more than one item share the same speakable text.
    \item \textit{Touch target} in Fig.~\ref{fig:items}(g) means clickable and long-clickable Views that are smaller than 48dp x 48dp in either dimension.
    \item \textit{Text contrast} in Fig.~\ref{fig:items}(h) means texts with a contrast ratio lower than 3.0 between the text color and background color.
    \item \textit{Image contrast} in Fig.~\ref{fig:items}(i) means images with a contrast ratio lower than 3.0 between the foreground and background color.
    \item \textit{Link} in Fig.~\ref{fig:items}(j) means URLSpan does not use an absolute URL. 
\end{itemize}

\section{Related Work}\label{sec:related}
In this section, we introduce related work on app accessibility testing and existing empirical studies on mobile app accessibility.

\subsection{Mobile Accessibility Testing}\label{subsec:access_approaches} 
Mobile apps have become a vital part of our day-to-day lives and are facing fierce competition. If the app is not easy to use (inaccessible), then users would probably abandon it and look for another app with similar functionality. On the other hand, for people with disabilities, the phenomenon is even more severe. Therefore, the accessibility testing to reduce accessibility problems in mobile apps is necessary and important. Although there has been research work investigating mobile apps testing methods~\cite{ monkey,mao2016sapienz,su2017guided}, mobile app accessibility testing is studied to a lesser extent. Informed by a recent survey study that provides an overview of available tools for detecting accessibility issues \cite{silva2018survey} and other related studies on accessibility testing \cite{daihua2015accessibility,lint}, we categorize accessibility testing related methods into two categories (i.e., static and dynamic mobile accessibility testing).
	
\subsubsection{\textbf{Static Accessibility Testing}}
Android Lint~\cite{lint} is a static code analyzer which is a part of Android Studio IDE~\cite{androidstudio}. It can report the errors such as missing translation, layout performance problems, and also accessibility problems like missing content descriptions. \textit{However, this method has been demonstrated to be ineffective for detecting mobile accessibility issues}~\cite{eler2018automated,fan2018efficiently,alshaybanaccessibility,fan2018large}. Other testing {tools} such as Espresso~\cite{web:androidEspresso} and Robolectric~\cite{web:androidRobolectric} can be used to detect accessibility issues. But these tools require developers to manually specify the testing cases and also embed the specific APIs into their apps which significantly increase developers' workload. Developers can also check the properties of GUI components after obtaining the layout of the user interface pages, or requires developers to interact with the accessibility tool to get the results. For example, the developers can use the screen reader (e.g., TalkBack~\cite{talkback} for Android, VoiceOver~\cite{voiceover} for iOS) to read the screen content and interact with their apps by certain gestures to check the app accessibility for users with vision impairment. They may also ask users with motor issues to check if they can easily reach all functionalities within the app. \textit{Although such manual exploration can mimic the real user experience, it is time-consuming and labor-intensive.} Apart from these static testing tools, some work focused on detecting specific types of accessibility issues (e.g., item label missing) by leveraging deep learning algorithms~\cite{chen2020unblind}.
		
\subsubsection{\textbf{Dynamic Accessibility Testing}}
Some tools are also released for assisting developers with accessibility testing via manual exploration of screens/UIs. Android UI Automator Viewer~\cite{uiautomatorviewer} provides a convenient GUI to scan and analyze the user interface components currently displayed on an Android device. Accessibility Scanner~\cite{scanner} is another tool released by Google for identifying accessibility issues within the current screen. \textit{However, the problem of these tools is that developers must activate the tool on the device in each screen of the app to get the results}~\cite{alshaybanaccessibility}. It means that it still requires manual exploration of the app, which is time-consuming and may also miss some functionalities of the apps (low activity coverage). That is also why few apps adopt these tools when developing their apps~\cite{vendome2019can}. 

To overcome the limitations of testing tools, Eler et al.~\cite{eler2018automated} developed a model to automatically generate testing cases specifically for accessibility testing. Similarly, to carry out a study of accessibility issues, Alshayban et al. leveraged the Android app testing tool, {Google Monkey}~\cite{monkey}, to explore the app screen to collect the accessibility issues. {Different from their work, our tool actually does not require test cases, inherits the results provided by Google {Accessibility Test Framework} for Android in which checking rules are developed by accessibility experts.}

\subsection{Empirical Studies of Mobile Accessibility}
Previous research investigating accessibility issues mainly focus on web applications~\cite{hackett2005retrospective,espadinha2011accessibility,Gilbertson:2012:GIM:2207016.2207024,billingham2014improving}. Recently, researchers have begun to investigate the accessibility issues of mobile apps in different domains, such as health~\cite{daihua2015accessibility}, public transportation \cite{Sanchez:2013:APB:2494091.2496002}, smart homes \cite{deOliveira:2016:ASH:3033701.3033730}, smart cities~\cite{smartcity}, and government engagement~\cite{goverment_engagement}. Kane et al.~\cite{kane2009freedom} carried out a study of mobile device adoption and accessibility for people with visual and motor disabilities. Ross et al.~\cite{ross2018examining} examined the image-based button labeling in a relatively larger number of android apps, and they specify some common labeling issues within the apps. In their further study~\cite{ross2020epidemiology}, they conducted their study from the perspective of accessibility issue types. They measured the prevalence of each accessibility issue across all relevant element classes (UI components) and apps. In other words, they focused on each issue type independently, which is a different research aspect compared with ours. Yan and Ramachandran~\cite{yan2019current} adopt the IBM Mobile Accessibility Checker to explore if 479 Android apps violate the accessibility guidelines and calculate the degree of violation. Vendome et al.~\cite{vendome2019can} observed the fact that developers rarely used accessibility APIs or assistive descriptions. They further create a taxonomy regarding the aspects of accessibility issues discussed by developers' posts on Stack Overflow. However, these works were based on the analysis of a relatively small number of mobile apps (no more than a few hundreds) instead of a large-scale dataset.

In the latest work, Alshaybana et al.~\cite{alshaybanaccessibility} conducted an empirical study on accessibility issues by leveraging the ability of Google Accessibility Test Framework~\cite{framework} and Google Monkey. For abbreviation, we call their study as \underline{A}ccessibility \underline{T}esting with \underline{Monkey} (\their) throughout the paper. From the apps perspective, they carried out a study at the screen level by using the criteria: {inaccessibility issue rate} for {{each page}}, and only investigated the distributions of inaccessibility issue rate for each app, each issue type, and app categories due to the limited issues (for each app) they collected using Monkey, such limitation is also acknowledged by them. Remarkably, the limited number of issues is enough for the prevalence of accessibility issues at the screen level, but difficult to carry out a more in-depth study at the issue level. As for the analysis from the apps perspective, they actually paid more attention to the analysis from the perspectives of developers and users instead of the accessibility issues themselves. In this paper, we aim to conduct an empirical investigation from the perspective of accessibility issues themselves and reveal more fine-grained findings compared with the existing studies. To this end, different from the previous works, we propose a fully automated and effective accessibility testing and issue collection tool with relatively high activity coverage to collect a large-scale and relatively comprehensive dataset of issues for this empirical investigation.

\section{App UI Exploration Tool}\label{sec:tool}
\begin{figure*}
	\centering
	\includegraphics[width=0.9\textwidth]{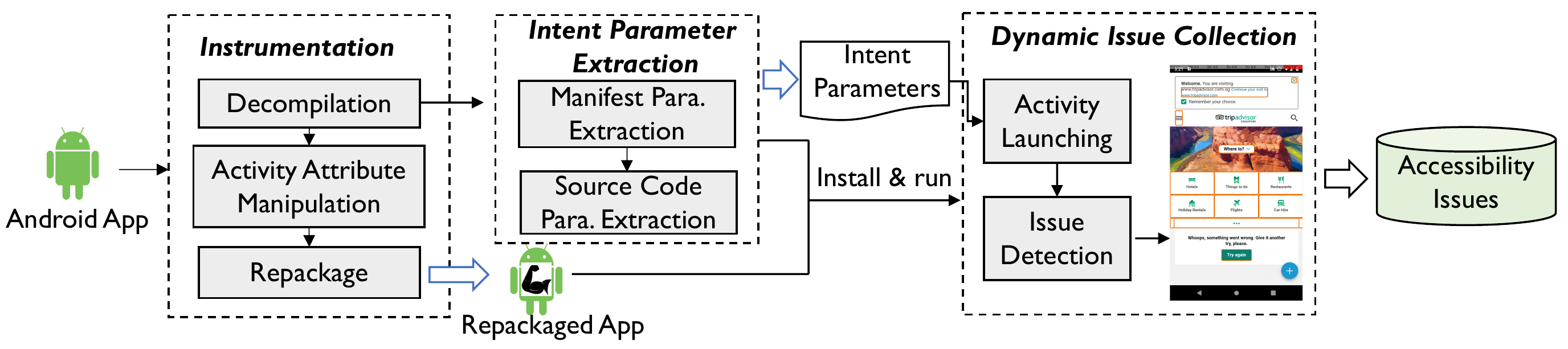}
	\caption{\rOne{Accessibility testing and issue collection with \ours}}
	\label{fig:overview}
\end{figure*}

To overcome the limitations of accessibility issue collection in the previous studies such as~\their, as shown in Fig.~\ref{fig:overview}, we propose a novel app UI exploration tool (named \textbf{\ours}) that can facilitate app accessibility testing and be used to collect issues effectively and efficiently. It leverages the instrumentation technique and static data-flow analysis based on \textit{Activity intent parameter extraction} to explore UI pages. Additionally, \ours integrates Google Accessibility Test Framework~\cite{framework} by feeding the explored app UI pages to it.

\subsection{\ours}\label{subsec:ours}
To capture the accessibility issues in app pages, we aim to automatically explore as many app screens as possible. Basically, dynamic app testing tools of Android apps such as Google Monkey~\cite{monkey}, Sapienz~\cite{mao2016sapienz}, and Stoat~\cite{su2017guided} are one choice to do this task, and Eler et al.~\cite{eler2018automated} and Alshayban et al.~\cite{alshaybanaccessibility} did it in this way. However, these tools are not suitable {enough} for accessibility testing of the app ecosystem due to the following aspects. (1) These app testing tools can only achieve around 40\% activity coverage (\S~\ref{subsubsec:monkey}), which is not satisfactory to check accessibility issues for apps. It would introduce data bias and it is difficult to show the real status of accessibility of apps. (2) It takes much more time for these testing tools to run each app. Such a task is time-consuming and labor-intensive. 
	
In fact, the core problem is to render or explore as many UI pages as possible. To our knowledge, two kinds of methods can be used to render UI pages: (1) Static page rendering, which can render the pages by using the static layout files (i.e., xml files) in the apk. However, according to a recent study~\cite{chen2019storydroid}, there are 62.3\% apps using dynamic layout method. Although Chen et al.~\cite{chen2019storydroid} proposed to transfer the dynamic layout types to static layout, the user interface differences between the generated pages and the original pages make accessibility analysis inaccurate. Therefore, we aim to render and explore app pages by dynamically loading the UI pages. (2) Dynamic page rendering, which can launch the pages by using Android \textit{adb}~\cite{andow2017uiref}, however, launching activities that require special fields (e.g., Intent parameters such as ``action'', ``category'', and Bundle data) would cause a crash with ``NullPointerException''. Such situation affects the accessibility testing and issue collection process. 
	
Specifically, as shown in Fig.~\ref{fig:overview}, \ours contains three main phases:
(1) app instrumentation, which instruments the apk files to enable launching by other third-party components; (2) activity intent parameter extraction, which extracts the required Activity {Intent} parameters for launching each activity; (3) accessibility issue collection, which dynamically launches pages and uses Google Accessibility Test Framework for further issue checking.

\begin{figure}
	\centering
	\includegraphics[width=0.475\textwidth]{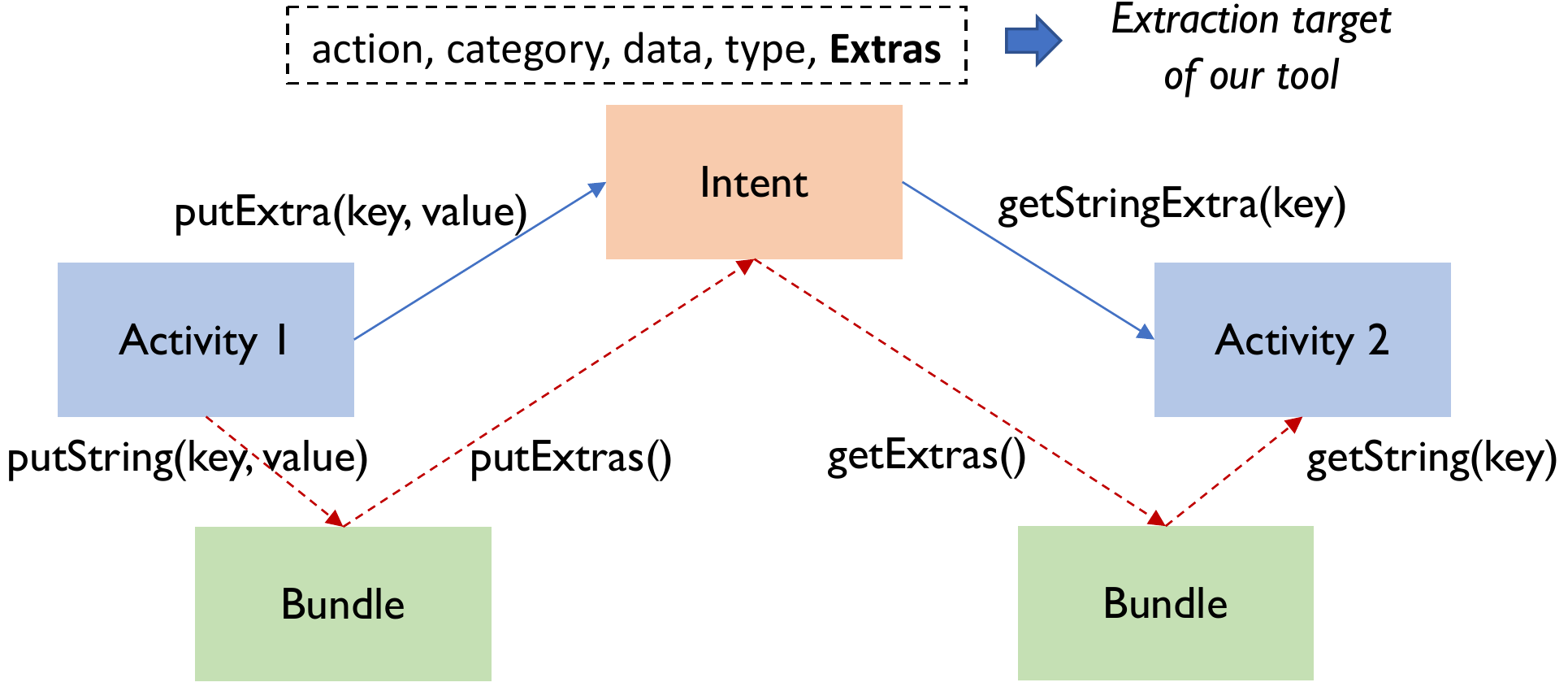}
	\caption{Data transfer between activities via {Intent}}
	\label{fig:intent}
\end{figure}

\subsubsection{{Instrumentation and Intent Parameter Extraction}}\label{subsubsec:two_phases}
To enable activity launching from other entries, we instrument each apk by manipulating the Android Manifest file (\sen{Activity Attribute Manipulation in Fig.~\ref{fig:overview}}) and repackage it to a new one. Specifically, \ours first decompiles the app (\sen{Decompilation in Fig.~\ref{fig:overview}}), extracts each activity together with its required fields such as ``action'', and sets the ``exported=true'' in order to enable the launching process from other components. We then repack it to a new apk (Repackage in Fig.~\ref{fig:overview}) and sign it to ensure the usability. Note that the repackaged apps are only used for experimental purpose, and all the experiments are conducted in a controlled environment. The repackaged apps will not be released for commercial use.

The second part (i.e., Activity Intent parameter extraction) is the core step of \ours, we leverage data-flow analysis to extract the Intent parameters required to launch the target activities. Fig.~\ref{fig:intent} shows the mechanism of activity launching, where Activity 1 puts data into the Intent object and sends it to Activity 2, and Activity 2 extracts the data out to render the UI pages. The parameters of Intent for launching Activity 2 are the extraction target of \ours, without them, Activity 2 may not be successfully launched. \ours is able to parse two categories of Intent parameters. As shown in Table~\ref{tbl:parameters},

\begin{table}
\centering
\caption{Types of Intent parameters}
\label{tbl:parameters}
\begin{tabular}{l|l|l}
\hline
\textbf{Type} & \multicolumn{2}{l}{\textbf{Sub-Type}} \\ \hline
\multirow{4}{*}{\begin{tabular}[c]{@{}l@{}}\textbf{Extracted Intent Parameters} \\ \textbf{ From} \textbf{Manifest File}\end{tabular}} & \multicolumn{2}{l}{Action} \\ \cline{2-3} 
 & \multicolumn{2}{l}{Category}  \\ \cline{2-3} 
 & \multicolumn{2}{l}{Data} \\ \cline{2-3} 
 & \multicolumn{2}{l}{Type}  \\ \hline
\multirow{5}{*}{\begin{tabular}[c]{@{}l@{}}\textbf{Extracted Intent Parameters} \\ \textbf{From}  \textbf{Source Code}\end{tabular}} & \multirow{5}{*}{Extras} & String  \\ \cline{3-3} 
 &  & Integer  \\ \cline{3-3} 
 &  & Long \\ \cline{3-3} 
 &  & Float\\ \cline{3-3} 
 &  & Boolean  \\ \hline
\end{tabular}
\end{table}
\begin{itemize}
    \item \textbf{a) Manifest Para. Extraction.} For the basic parameters such as action, category, data, and type, we parse them from the Android Manifest file and record the mapping relations between activities and these basic parameters.
    \item \textbf{b) Source Code Para. Extraction.} For the Intent extras parameters, we extract them from source code through data-flow analysis. We consider extracting two types of Intent data described as follows.
\end{itemize}
 
One data type is transferred from ``Activity1'' to ``Activity2'' by using Intent directly. The data passing step is ``create an Intent object''$\rightarrow$``call intent.putExtra''$\rightarrow$``call StartActivity(intent) to pass the Intent''$\rightarrow$``call intent.getStringExtra'' to get the transferred data (the blue flow demonstrated in Fig.~\ref{fig:intent}). The other data type uses {Bundle} mechanism to transfer a bundle of data from ``Activity1'' to ``Activity2''. The data passing step is ``create Intent and {Bundle} objects''$\rightarrow$``call bundle.putString and intent.putExtras(bundle)''$\rightarrow$``call StartActivity(intent) to pass the Intent''$\rightarrow$``call getIntent().getExtras and bundle.getString'' to get the transferred data (the red flow demonstrated in Fig.~\ref{fig:intent}). As shown in Algorithm~\ref{algo:parameter}, for the Intent parameters extraction, we first obtain the basic parameters from manifest file (Line 3). We then filter the methods related to activity life cycle (Line 5), called $meths$. These life cycle methods like \textit{onCreate()} and \textit{onStart()} contain Intent extras parameters for rendering app pages. For each $meths$, if it calls specific APIs like \textit{getStringExtra} and \textit{getExtras} (Line 7), we trace the parameters' key through backward data-flow analysis (Line 8). Note that, the value type of each parameter is based on the corresponding API. The Intent extras parameters may not be obtained in life cycle method. For these cases, we trace the callee method (Line 10) by parsing the call graph and then extract the parameters through the same way for life cycle method (Line 11). After that, we can obtain the Intent extras parameters $paras\_extras$ for further accessibility 
testing of rendered pages.

\begin{algorithm2e}[t]\small
\setcounter{AlgoLine}{0}
\caption{Intent Parameters Extraction}
\label{algo:parameter}
\DontPrintSemicolon
\SetCommentSty{mycommfont}
\KwIn{$apk$\;
\KwOut{$paras\_intent$}
$all\_acts$ $\leftarrow$ getAllActivities($all\_classes$) \;
$cg$ $\leftarrow$ getCallGraph($apk$) \;
$paras\_basic$ $\leftarrow$ getBasicIntentParameters($manifest$) \;
\ForEach{$act$ $\in$ $all\_acts$}{
	$meths$ $\leftarrow$ getLifeCycleCallBacks($act$) \;
	\ForEach{$m$ $\in$ $meths$}{
	    \If{hasExtrasParameters($m$)}{
	        $para\_extras$ $\gets$ backwardDataFlowAnalysis($m$)\;
	    }
	    \Else{
	        $m\_callee$ $\gets$ getCallerMethod($m$, $cg$)\;
	        $para\_extras$ $\gets$ getExtrasParameters($m\_callee$)\;
	    }
	}
}
\Return $paras\_intent$ $\gets$ $paras\_basic$ $\bigcup$ $paras\_extras$ \; 
}
\end{algorithm2e}

\subsubsection{{Accessibility Testing with \ours and Issue Collection}}\label{subsec:collection}
To dynamically launch each activity, \sen{as shown in Fig.~\ref{fig:overview}}, we install the new repackaged apk on the Android emulator, and attach the Intent parameters extracted by our tool to the current activity. When it is launched successfully (Activity Launching in Fig.~\ref{fig:overview}), we take screenshots of each app page and then feed it to Google Accessibility Test Framework~\cite{framework}. Meanwhile, for activities that fail to launch due to app crashes or permission required, we dump the layout hierarchy of the current activity and analyze it to check whether it contains keywords (e.g., ``has stopped'' and ``keeps stopping'' for app crash, ``ALLOW'' and ``DENY'' for permission required), and grant the permission required to proceed. When the app crashes, we stop the app and set it to the original state (i.e., a fresh state for another activity to launch). We collect the detected accessibility issues (Issue Detection in Fig.~\ref{fig:overview}) and the corresponding layout hierarchy of each page that contains accessibility issues.

\subsection{RQ1: Evaluation of \ours}\label{subsec:too_evaluation}
In this section, we evaluate the effectiveness and efficiency of \ours by comparing it with manual exploration and Monkey. We mainly compare the explored activities coverage and the time cost since both tools rely on the same accessibility test framework to check accessibility issues, the main difference comes from the number of explored activities.

\subsubsection{{Manual exploration with Google Scanner vs. \ours}}\label{subsubsec:manual}
We conduct a user study to compare \ours with manual exploration. We recruit 10 participants from our university, including Ph.D students, post doctorates, and undergraduate students. We randomly select four apps (i.e., \textit{Bitcoin}~\cite{bitcoin}, \textit{Bankdroid}~\cite{bankdroid}, \textit{ConnectBot}~\cite{connectbot}, and \textit{Vespucci}~\cite{vespucci}) from Google Play Store, and ask them to use {Accessibility Scanner} to detect accessibility issues on these four apps in a fixed time (i.e., 10 minutes per app), trying to explore as many pages as possible, meanwhile, we record the number of collected issues. In contrast, we use \ours on these four apps to detect accessibility issues, and record the time and the number of detected issues. As shown in Table~\ref{tbl:efficient}, the result shows that the participants can only explore 40.80\% user interface pages for each app on average, \sen{collecting 79 accessibility issues}. While \ours explores 91.84\% pages per app on average, and collects 142 accessibility issues in total. Moreover, it only takes 2.65 minutes for \ours to test one app, and it is about 4 times (10 mins) faster than that of manual exploration. To understand the significance of the differences between manual exploration and with \ours, we carry out the Mann-Whitney U test~\cite{utest}, which is designed for small samples. Table~\ref{tbl:efficient} shows that our result is significant with p-value $<$ 0.01. Obviously, \ours is significantly more effective and efficient in collecting accessibility issues, and can help developers explore more pages, increasing the possibility of detecting more potential accessibility issues.

\begin{table}\footnotesize
\centering
\caption{Effectiveness and Efficiency Evaluation of \ours}
\label{tbl:efficient}
\scalebox{0.9}{\begin{tabular}{l|c|c||c|c}
\hline
\textbf{Metrics} & \begin{tabular}[c]{@{}c@{}}\textbf{Manual} \\\textbf{Exploration}\end{tabular} &
\begin{tabular}[c]{@{}c@{}}\textbf{\ours} \end{tabular} &
\begin{tabular}[c]{@{}c@{}}\textbf{Monkey}\end{tabular} & \begin{tabular}[c]{@{}c@{}}\textbf{\ours} \end{tabular} \\ \hline
\textbf{Avg Time (min)} & 10 & \revised{2.65} & 30 & \revised{5.67} \\ \hline
\begin{tabular}[c]{@{}l@{}}\textbf{Avg launched} \\\textbf{Activity Ratio}\end{tabular} & \revised{40.80\%} & \revised{91.84\%} & 43.09\% & \revised{79.81\%} \\ \hline
\begin{tabular}[c]{@{}l@{}}\textbf{\#Collected Issues}\end{tabular} & \changed{79} & \changed{142} & \changed{851} & \changed{3,063} \\ \hline
\end{tabular}}
\begin{center}
    \textit{The number of apps for manual testing and testing with Monkey are 4 and 100, respectively.}
\end{center}
\end{table}

\begin{figure}
	\centering
	\includegraphics[width=0.325\textwidth]{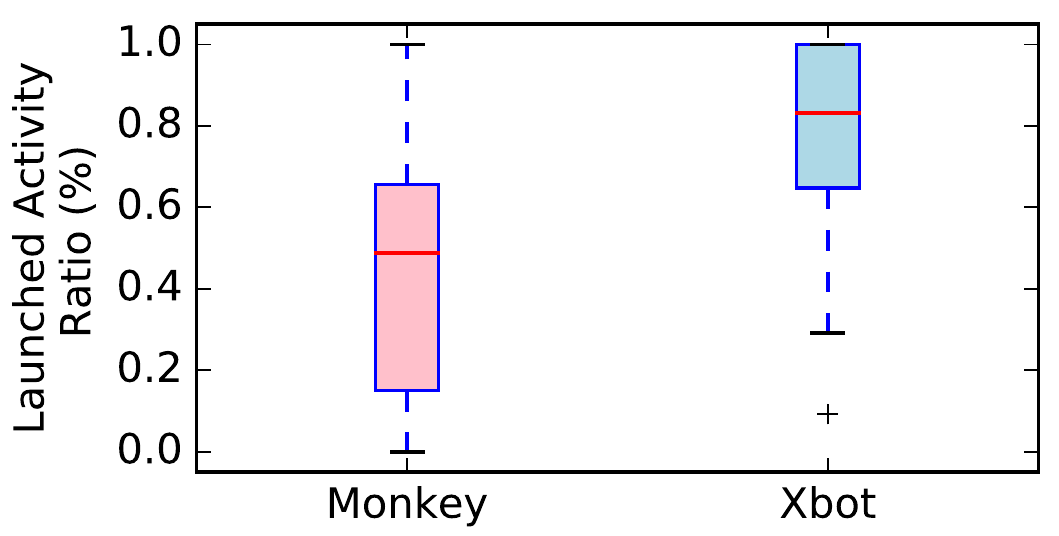}
	\caption{Comparison on activity coverage of Monkey and \ours for accessibility testing}
	\label{fig:compare}
\end{figure}

\subsubsection{\revised{Accessibility testing with Google Monkey vs. \ours}}\label{subsubsec:monkey}
Besides the manual exploration method with Accessibility Scanner, using dynamic Android app testing tools such as Google Monkey is another method for automated accessibility testing in previous work~\cite{eler2018automated, alshaybanaccessibility}. To demonstrate the better performance of \ours, we choose the most representative Android app testing tool, {Monkey}~\cite{monkey}, which is also the official testing tool of Google and widely-used in both academy and industry. Specifically, we randomly collect 50 commercial apps from Google Play and 50 open-source apps from F-Droid~\cite{f-droid} as the experiment subjects. For the dynamic exploration with Monkey, we configure the execution parameter as ``--ignore-crashes --ignore-timeouts --throttle 250 -v -v -v 50000''. The parameter configuration means that Monkey will ignore crashes and timeouts and the time interval between two events is 250 ms. The execution time is set by 30 minutes and the experiment environment is the same as \ours mentioned in \S~\ref{subsec:ours}. Fig.~\ref{fig:compare} shows the comparison result, the average launched activity ratios of 100 Android apps are 43.09\% vs. 79.81\% for the two methods. 
\sen{In terms of the difference of collected accessibility issues between Xbot and the collection method by using Monkey, Xbot is able to collect 3 more times (3,063 vs. 851) accessibility issues. The results unveil that
\ours outperforms Monkey when checking and collecting accessibility issues dynamically.} As shown in Fig.~\ref{fig:compare}, we can see that the launched activity ratio of testing with Monkey ranges from \textbf{15\% to 65\%}. \ours performs better and the average launched activity ratio of testing is about \textbf{80\%}. We also conduct a statistic analysis for their ability of activity launching in mobile accessibility testing, the p-value $<$ 0.01, which means that the results of these two methods are significantly different.

\begin{table}\normalsize
\centering
\caption{\sen{Comparison between \ours and \their on issue collection}}
\label{tbl:comparewithicse}
\sen{\scalebox{0.9}{\begin{tabular}{l|c}
\hline
\textbf{Method} &
\begin{tabular}[c]{@{}c@{}}\textbf{\#Collected Issues}\end{tabular} \\ \hline
\begin{tabular}[c]{@{}l@{}}\textbf{\their \cite{alshaybanaccessibility}}\end{tabular} & \changed{9,462} \\ \hline 
\begin{tabular}[c]{@{}l@{}}\textbf{\ours}\end{tabular} & \changed{63,734} \\ \hline 
\end{tabular}}} 
\begin{center}
    \footnotesize
    \textit{\sen{The comparison is based on the dataset in \their \cite{alshaybanaccessibility}.}}
\end{center}
\end{table}

Besides the above basic evaluation, to conduct a fair comparison, we also evaluate the performance of Xbot by comparing it with \their's method~\cite{alshaybanaccessibility} in terms of issue collection on their released dataset~\cite{atmonkey-github}. We run Xbot and their tool \cite{atmonkey-github} on their dataset individually to explore the app UI pages and then collect the corresponding accessibility issues. As shown in Table~\ref{tbl:comparewithicse}, in terms of the number of collected accessibility issues, we are able to collect more issues obviously (\changed{63,734 vs. 9,462} on \their's dataset), owing to the effectiveness of \ours. The result is consistent with the result in the above evaluation on 100 Android apps.

\smallskip
\noindent\fbox{
\parbox{0.95\linewidth}{
\noindent \textbf{Answer to RQ1.} \ours outperforms existing methods when conducting accessibility testing for Android apps. \sen{With the ability of app UI exploration with relatively high activity coverage (about 80\%), \ours is able to collect a \sen{relatively comprehensive} and large-scale dataset of accessibility issues effectively and efficiently for further empirical investigation at the issue level}.}}

\section{\sen{Empirical Investigation of App Accessibility}}\label{sec:new_study}
In this section, we \revised{aim to} conduct \revised{an empirical study} on the large-scale dataset collected by \ours to mine the accessibility issue characteristics. \sen{Therefore, we pay more attention to the analysis from the perspective of accessibility issues themselves in this paper.}
(1) We first {investigate} the current status quo of the accessibility issues in apps including both the prevalence and severity situation at the issue level. 
(2) Then, we mine the in-depth relation between issue types and app categories, GUI component types.
(3) Thirdly, as we conducted quantitative analysis on specific issue types while \cite{alshaybanaccessibility,ross2020epidemiology} do not, we can provide more quantitative issue details and more fine-grained findings for app developers. 
(4) Last, we further analyze the fixing status using our collected dataset 
and discussed the tracking result in \their~\cite{alshaybanaccessibility}.

\begin{table}\footnotesize
	\caption{Accessibility issues collected by \ours and the corresponding features. (W.: With; Lau.: Launched)}
	\label{tbl:basic}
	\scalebox{0.88}{\begin{tabular}{cccccc||c}
		\hline
		{\bf Source} &
		\textbf{\#Apps}
		& \begin{tabular}[c]{@{}c@{}}{\bf \#Apps W.} \\ {\bf Issue(s)}\end{tabular} & {\bf \#Acts} & \begin{tabular}[c]{@{}c@{}}{\bf \#Lau.}\\ {\bf Acts}\end{tabular} & \begin{tabular}[c]{@{}c@{}}{\bf \#Acts W.}\\\ {\bf Issue(s)}\end{tabular} &  {\bf \#Issues} \\ \hline
		{\begin{tabular}[c]{@{}c@{}}{\bf Google}\\\ {\bf Play}\end{tabular}} & 1,172 & \cellcolor{grayone}{
		\begin{tabular}[c]{@{}c@{}}{\bf 1,082}\\\ {\bf (92.32\%)}\end{tabular}} & 17,926 & \cellcolor{grayone}{\begin{tabular}[c]{@{}c@{}}{\bf 12,685}\\\ {\bf (70.76\%)}\end{tabular}} & \cellcolor{grayone}{\begin{tabular}[c]{@{}c@{}}{\bf 10,298}\\\ {\bf (81.18\%)}\end{tabular}}  & 66,687 \\ \hline
		{\bf F-Droid} & 1,098 & \cellcolor{grayone}{ \begin{tabular}[c]{@{}c@{}}{\bf 938}\\\ {\bf (85.42\%)}\end{tabular}} & 5,995 & \cellcolor{grayone}{\begin{tabular}[c]{@{}c@{}}{\bf 4,732}\\\ {\bf (78.93\%)}\end{tabular}}& \cellcolor{grayone}{ \begin{tabular}[c]{@{}c@{}}{\bf 3,079}\\\ {\bf (65.07\%)}\end{tabular}} & 20,080 \\ \hline
		{\bf Total} & 2,270 & \cellcolor{graytwo}{\begin{tabular}[c]{@{}c@{}}{\bf 2,020}\\\ {\bf (88.99\%)}\end{tabular}} & 23,921 & \cellcolor{graytwo}{\begin{tabular}[c]{@{}c@{}}{\bf 17,417}\\\ {\bf (72.81\%)}\end{tabular}} & \cellcolor{graytwo}{\begin{tabular}[c]{@{}c@{}}{\bf 13,377}\\\ {\bf (76.80\%)}\end{tabular}} & \cellcolor{graytwo}{\bf \textit{86,767}} \\ \hline
	\end{tabular}}
\end{table}

Table~\ref{tbl:basic} summarizes all related data that we use to quantitatively analyze the app accessibility issues, including the accessibility issues collected by \ours. We execute 2,270 unique Android apps by \ours, including 1,172 closed-source apps from Google Play Store and 1,098 open-source apps from F-Droid. Since some apps may be available on both Google Play and F-Droid, we consider such apps as open-source apps to ensure there is no overlap and avoid biased results on closed-source vs. open-source apps. These apps contain 23,921 activities, and the activity coverage of \ours is 72.81\% (i.e., $\frac{\#Launched\,acts}{\#Acts}$), which is lower than the result of the average coverage for each app (i.e., 79.81\%) in \S~\ref{subsubsec:monkey}. Because some apps contain hundreds of activities, which largely affects \#Launched\,acts. Overall, \ours achieves a higher activity coverage on F-Droid apps than Google Play apps (i.e., 78.93\% vs. 70.76\%).

\subsection{RQ2: Overall Status of Mobile App Accessibility}\label{subsec:status}
Among the 2,270 apps, we finally collect 86,767 real accessibility issues in total, which is the largest dataset so far in this research area.\footnote{Besides the 86,767 accessibility issues, we also obtain other 63,734 issues collected from the evaluation of Xbot (RQ1). Therefore, we actually have over 100k accessibility issues in total.} 2,020 (88.99\%) Android apps in our dataset contain at least one accessibility issue. This result demonstrates that accessibility issues are prevalent across all apps (prevalence situation), which is consistent with the {conclusion} drawn by Alshayban et al.~\cite{alshaybanaccessibility}. However, they only revealed the prevalence of issues at the screen level due to the limited number of issues collected for each app, while we further provide an empirical investigation of the overall status of app accessibility at the issue level to show the severity situation as follows. We use the number of issues on each UI page and in each app to reflect the severity situation. Specifically, on average, there are 43 accessibility issues for each app (i.e., $\frac{\#Issues}{\#Apps\,with\,issue(s)}$). Among the 17,417 launched activities, there are 6.5 accessibility issues on average for each flawed page (i.e., $\frac{\#Issues}{\#Acts\,with\,issue(s)}$).

We further investigate the differences of app accessibility between the closed-source and open-source apps, which is not investigated in the previous studies. Out of our expectation, compared with open-source apps, the commercial apps have a higher ratio (i.e., $\frac{\#Acts\,with\, issue(s)}{\#Launched\,acts}$) (65.07\% vs. 81.18\%) of accessibility issues. It identifies that the developers and the corresponding commercial companies do not pay sufficient attention to the accessibility issues in practice. On the other hand, although it seems that open-source apps are more accessible, that is because the open-source apps may have fewer features, i.e., fewer components in each page, leading to fewer accessibility issues. Specifically, each F-Droid app contains 5.5 activities, and each Google Play app contains 15.3 activities on average (i.e., $\frac{\#Acts}{\#Unique\,apps}$).

\smallskip
\noindent\fbox{
\parbox{0.95\linewidth}{
\noindent {\textbf{Answer to RQ2.}} 89\% apps in our dataset are suffering from accessibility issues, with 43 issues for each app and 6.5 issues for each page on average. Overall, open-source apps have a better status than closed-source apps in our dataset. The app accessibility deserves more attention from the development team.}}

\subsection{\sen{RQ3: In-depth Relation between Issue Type and App Category, GUI Component}}\label{subsec:cross_analysis}
\subsubsection{Accessibility issue types}
In this section, we conduct cross analysis of issue types vs. app category and GUI component (i.e., how frequently do issue types occur in various app categories, and in various GUI components), \sen{which has never been investigated in the previous studies~\cite{alshaybanaccessibility, ross2020epidemiology}}.

\begin{figure}
	\centering
	\includegraphics[width=0.425\textwidth]{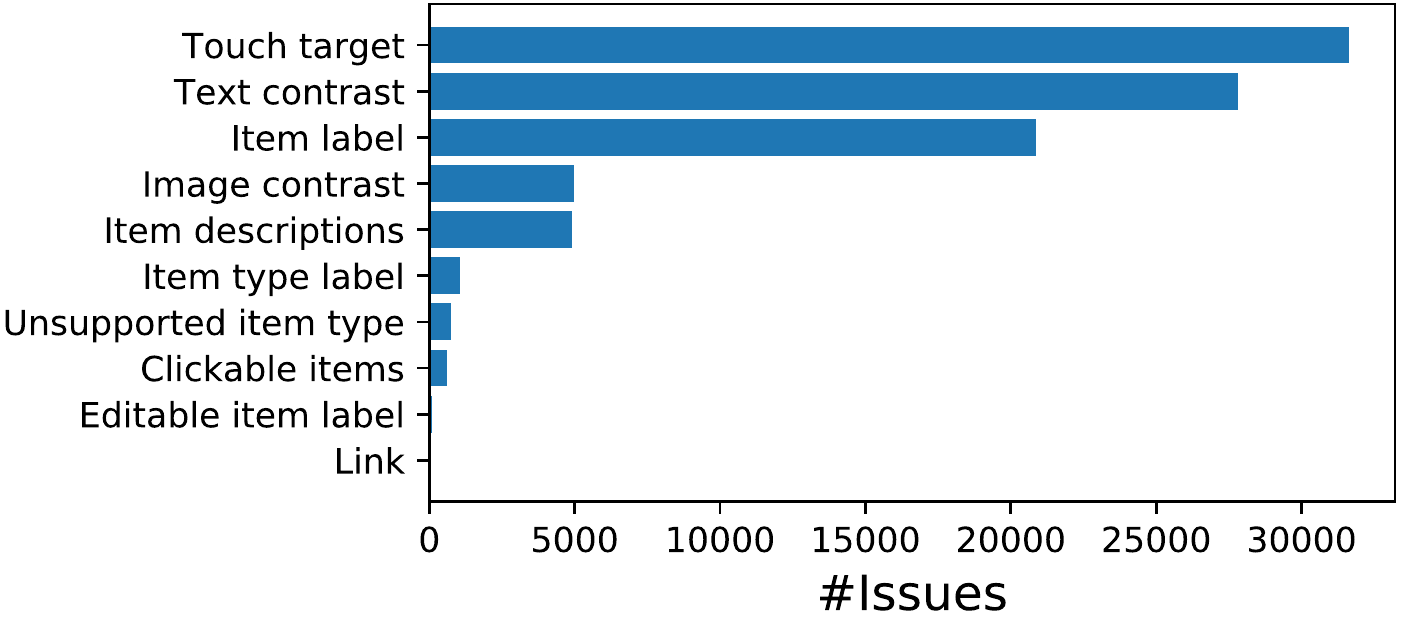}
	\caption{Issue type distribution ranked by \#issues}
	\label{fig:distribution}
\end{figure}

Specifically, 
to analyze the common accessibility issue types regarding app categories and GUI component types at the issue level, we firstly investigate the issue type distribution ranked by the number of accessibility issues. As shown in Fig.~\ref{fig:distribution}, \textit{item label}, \textit{item descriptions}, \textit{touch target}, \textit{text contrast}, and \textit{image contrast} are much more frequent compared with other accessibility issue types, accounting for {93.1\%} of all issues. They pose a serious challenge to the accessibility of user experience in apps and developers should pay more attention to them. Among them, \textit{touch target}, \textit{text contrast}, and \textit{item label} are the top 3 issue types ranked by the number of accessibility issues. These three issue types all contain over 20,000 issues. Compared with our study, Alshayban et al.~\cite{alshaybanaccessibility} only focused on the relations between issue types and apps, app categories based on the metric of inaccessibility issue rate at the screen level, while the in-depth relation between issue type and app category, GUI component at the screen level is not investigated in their study.

\subsubsection{{Different issue types in each app category}}\label{subsec:app_category}
\begin{figure}
	\centering
	\includegraphics[width=0.45\textwidth]{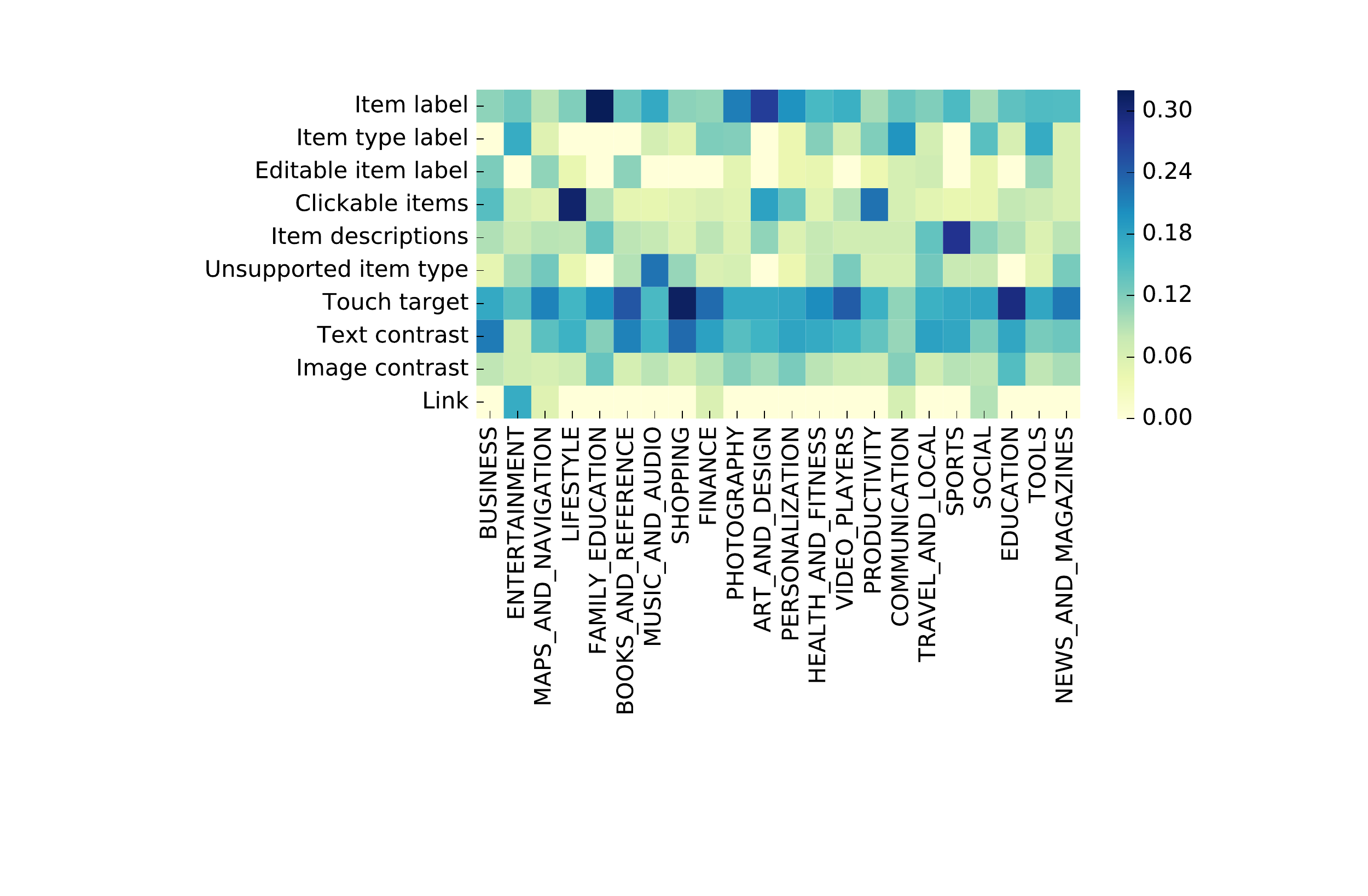}
	\caption{Different accessibility issues in different app categories (Issues in each app category are normalized to 1)}
	\label{fig:category_issue}
\end{figure}

To explore what types of accessibility issues often cause in different app categories, we compute the relative frequency of different types of issues within each app category. We draw a heat map in Fig.~\ref{fig:category_issue}, and the degree of the color in each cell represents the proportion of all issue types in each app categories. Within each column, the total number of 10 issue types add up to 1 and the darker color indicates the more issues of that type in this app category. We can see that some issues widely appear in most categories such as \textit{item label}, \textit{touch target}, and \textit{text contrast}, while some issues like \textit{editable item label}, \textit{link} rarely appear. 
	
On the other hand, some issues are rather severe in some categories than others. In other words, some specific app categories are more likely to have specific types of issues according to the relation between issue type and app category. For example, \textit{touch target} is a common issue for most app categories, but it is particularly serious for shopping apps. Shopping apps often offer their users a list of products to choose from per screen page. To accommodate so many elements within each page, they make the buttons too small which may cause difficulty for users to click them especially for the elderly. Similarly, \textit{Item descriptions} often occurs in sports app. Most sports apps are providing sports news, match living for users. To give users an overview of the team ranking, or broadcast list, they need to put many items in one page. Adding descriptions to each item is always difficult, especially that most lists are dynamically updated. For saving efforts, many developers just put the same content description (similar to alt text of the picture in the image~\cite{alt}) to all of these items like ``game'', ``video''. However, these identical descriptions for different items will confuse blind users who rely on the screen reader to read the content in the app.

\subsubsection{{Issue types related to GUI component types}}\label{subsubsec:component}
Within each flawed screen, the existence of issues is also highly related to the GUI components types such as \textit{TextView}, \textit{ImageView}, and \textit{Button}. 93.1\% accessibility issues belong to 5 components (i.e., \textit{TextView}, \textit{ImageView}, \textit{Button}, \textit{EditText}, and \textit{ImageButton}). Although some types of components such as \textit{TextInputLayout} and \textit{RadioButton} are not used frequently in apps, the issue percentage is very high (i.e., 65.8\% and 47.5\%). It means that designers and developers are more likely to make mistakes about accessibility when developing these specific components. These components deserve more attention from the development team.

\begin{figure}
	\centering
	\includegraphics[width=0.475\textwidth]{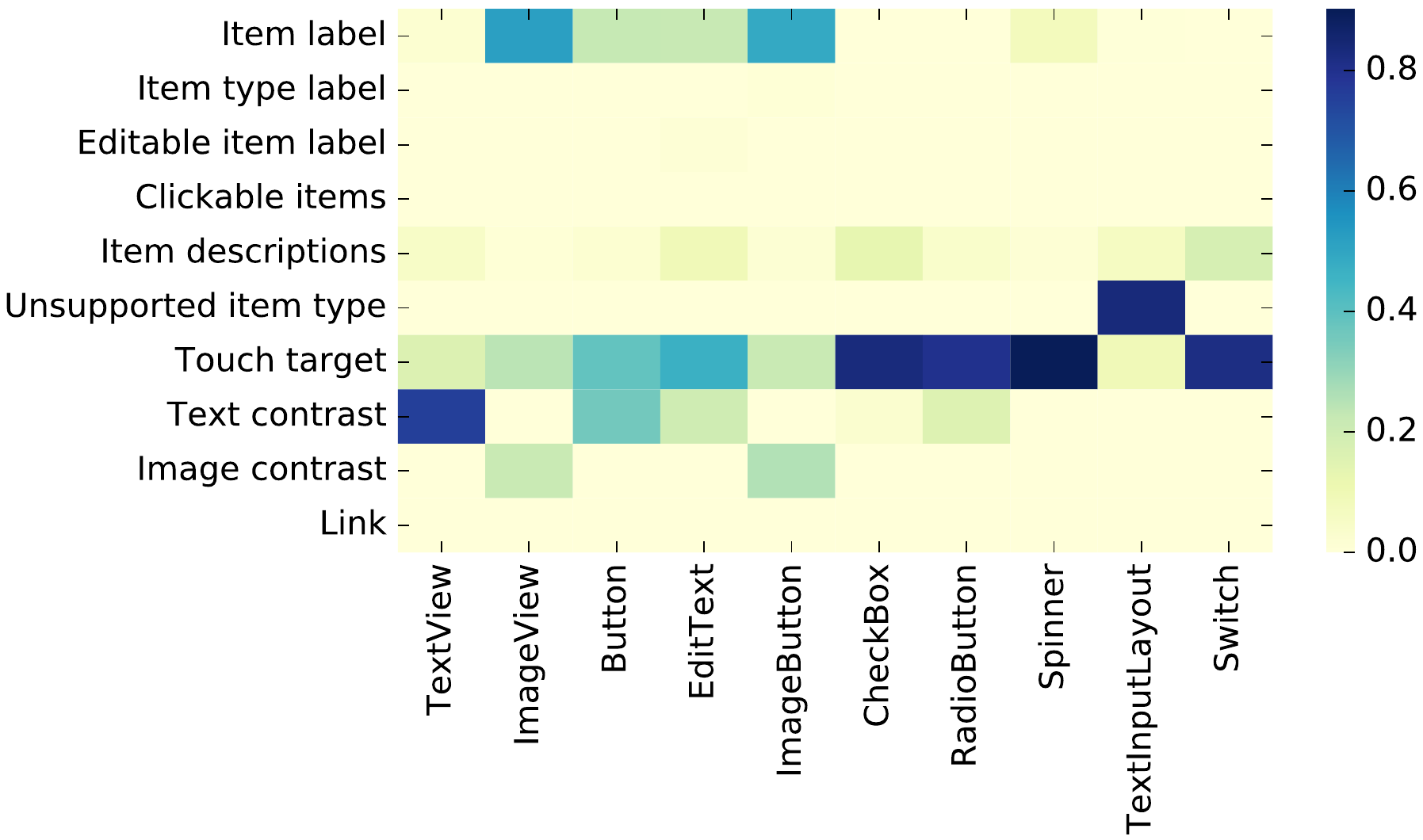}
	\caption{Different accessibility issues in different components (Issues in each component type are normalized to 1)}
	\label{fig:compt_issuetype}
\end{figure}
	
Some types of issues are also specifically related to certain components. To investigate their relation, we compute the percentage of different types of accessibility issues for each component type, and draw a heat map in Fig.~\ref{fig:compt_issuetype}. The issue \textit{touch target} frequently appears in clickable components such as \textit{Checkbox}, \textit{RadioButton}, \textit{Spinner}, and \textit{Switch}, as these components may be too small to be clicked by the users, especially for users with motion disability. 38.8\% accessibility issues of \textit{TextView} are about \textit{text contrast} issues which makes the content difficult to be read by users. For image-related components like \textit{ImageView} and \textit{ImageButton}, the biggest issue is the \textit{item label}, i.e., missing the content description of the image for users who cannot see the screen.
	
\smallskip
\noindent\fbox{
\parbox{0.95\linewidth}{
\noindent \textbf{Answer to RQ3.} 5 types (e.g., {touch target}, {text contrast}, and item label) of issues occur frequently. Some issue types are highly related to app categories such as the small size of touchable components in shopping apps and duplicate content descriptions of different items in sports apps. Similar patterns also apply to different component types such as the low text contrast in {TextView} and missing labels for image based GUI components.}}

\subsection{RQ4: Quantitative Analysis of Specific Issue Types}\label{subsec:specific_issues}
Based on the results in \S~\ref{subsec:cross_analysis}, we find that some issue types are more frequent and common than others such as \textit{text contrast}, \textit{image contrast} which are about the color contrast, and \textit{touch target} which is about the size of the component. In this section, we further provide an in-depth analysis on these three most frequent issue types.

\begin{figure}
	\centering
	\includegraphics[width=0.5\textwidth]{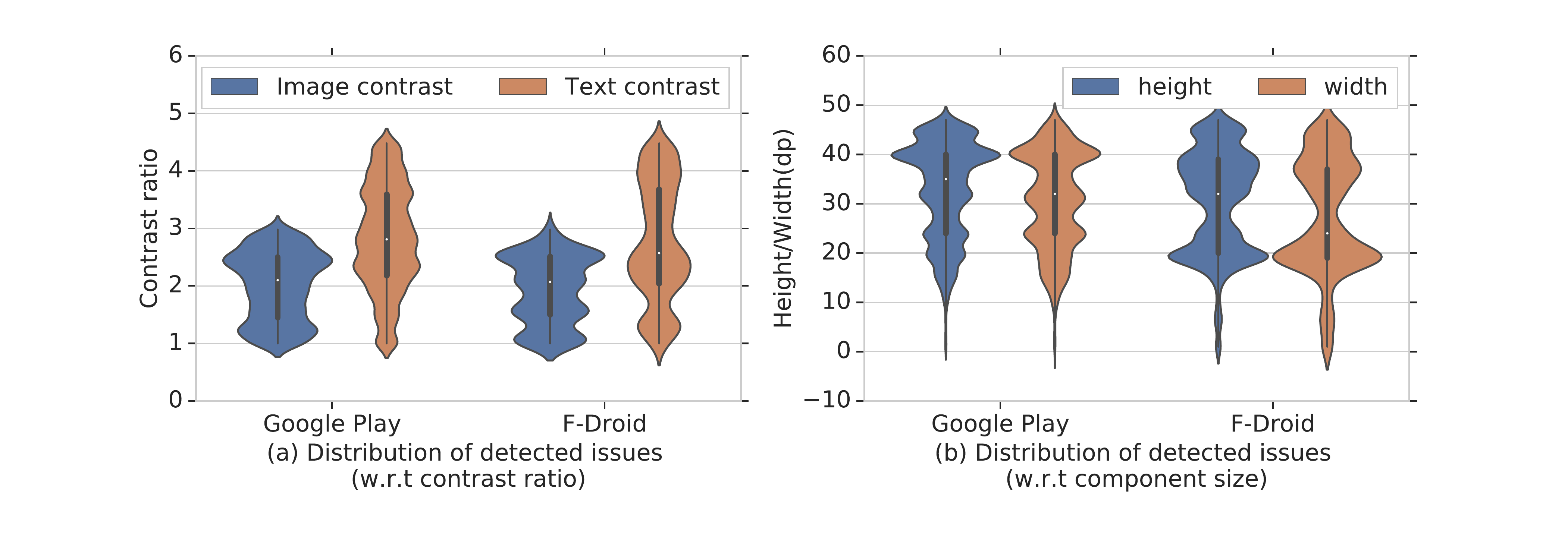}
	\caption{Distribution of the specific issue types (i.e., contrast ratio and component size of touch target issues)}
	\label{fig:quantitative}
\end{figure}

\begin{figure}
	\centering
	\includegraphics[width=0.5\textwidth]{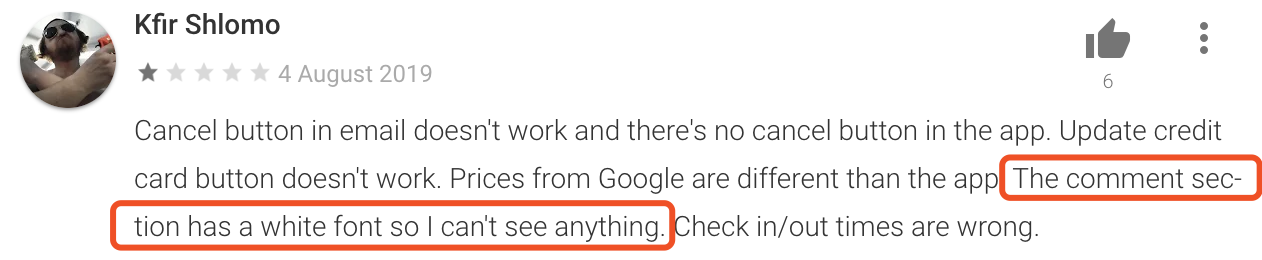}
	\caption{A real review complaining about text contrast}
	\label{fig:reviews}
\end{figure}

\begin{figure*}
		\centering
		\includegraphics[width=0.95\textwidth]{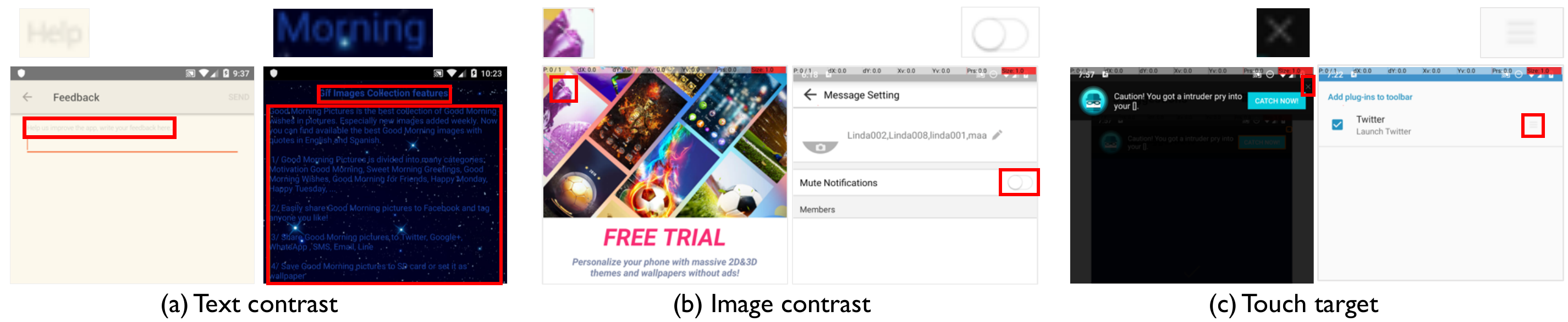}
		\caption{Real examples of accessibility issues of \textit{text contrast}, \textit{image contrast}, \textit{touch target}, \textit{item label}, and \textit{item descriptions}} 
		\label{fig:color_example}
\end{figure*}

The text contrast is the difference between the foreground text and the background color. Fig.~\ref{fig:quantitative} (a) shows that the overall results of the wrong text contrast ratio between Google Play and F-Droid are similar, ranging from 1 to 4.5 roughly. Most wrong instances are located between 2 to 4 contrast ratio, though the best practice of text contrast ratio is over 4.5 (including 4.5). We list the top-10 most frequent wrong pairs of foreground text and background color in Table~\ref{tbl:color} including gray text in white background, white text in gray background, blue text in red background (i.e., \#B05656). These color pairs will negatively influence the readability of the text, resulting in bad user experience. As shown in Fig.~\ref{fig:reviews}, the user named ``Kfir Shlomo'' complained ``The comment section has a white font so I cannot see anything.'' which is due to the accessibility issue of \textit{text contrast}. It is hard even for users without disabilities to discriminate the text from the background color, let alone the users with vision impairment or color blind~\cite{rello2012optimal}. More examples can be seen in the first two sub-figures in Fig.~\ref{fig:color_example} (a) and (b).
	
Compared with the results on text contrast issues, the results of image contrast also have a similar presentation for these two markets. Specifically, compared with Google Play apps, F-Droid apps have a wide range contrast ratio from 1 to 3. There are several cases that have a significant effect on a lower image contrast (i.e., around 1) for both two markets, which are also far away from the best practice of image contrast ratio. In addition, the contrast range between 2 and 3 accounts for the most image contrast issues for both two markets. As shown in Fig.~\ref{fig:color_example} (c), the item size is too small to see clearly for end-users, even for users without any disability. Some of small-size buttons are created intentionally regardless of the app accessibility. For example,  the ``close button'' in the left figure in Fig.~\ref{fig:color_example} (c) is so small that users have a great chance of clicking the ``CATCH NOW!'' button i.e., the advertisement.
	
\begin{table}
\small
\centering
\caption{Demo of the top 10 contrast issues}
\begin{tabular}{cccc}
	\hline
	{\bf Contrast Demo} & {\bf Foreground} & {\bf Background} & {\bf \#Issues} \\ \hline
	\includegraphics[width=0.04\textwidth]{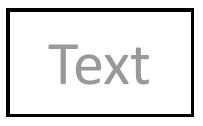}	& \#999999 & \#FFFFFF & 458 \\ \hline
	\includegraphics[width=0.04\textwidth]{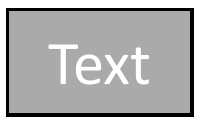}	& \#FFFFFF & \#{\footnotesize AAAAAA} & 388 \\ \hline
	\includegraphics[width=0.04\textwidth]{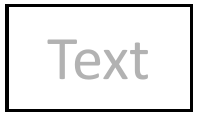}	& \#B2B2B2 & \#FFFFFF & 357 \\ \hline
	\includegraphics[width=0.04\textwidth]{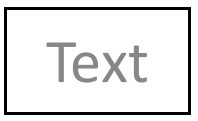}	& \#878787 & \#FFFFFF & 239 \\ \hline
	\includegraphics[width=0.04\textwidth]{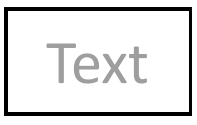}	& \#9E9E9E & \#FFFFFF & 230 \\ \hline
	\includegraphics[width=0.04\textwidth]{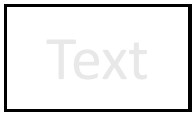}	& \#E8E8E8 & \#FFFFFF & 222 \\ \hline
	\includegraphics[width=0.04\textwidth]{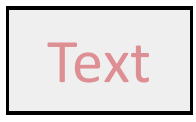}	& \#DE8F94 & \#EFEFEF & 217 \\ \hline
	\includegraphics[width=0.04\textwidth]{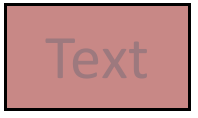}	& \#9D797E & \#C88886 & 217 \\ \hline
	\includegraphics[width=0.04\textwidth]{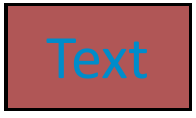}	& \#008CCA & \#B05656 & 212 \\ \hline
	\includegraphics[width=0.04\textwidth]{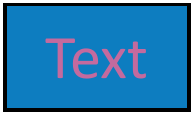}	& \#C46A9E & \#7755CD & 196 \\ \hline
\end{tabular}
\label{tbl:color}
\end{table}

Fig.~\ref{fig:quantitative} (b) summarizes the distribution of concrete component size in detected touch target issues. The distributions of component width in terms of Google Play and F-Droid are different obviously. Specifically, the component width distribution of Google Play is mainly ranging from 20dp to 40dp, however, the distribution of F-Droid is very concentrated on 20dp. While the best practice of the height and width is larger than 48 dp. In other words, there are strong commonalities for such issues in F-Droid apps, meanwhile, their touch target components in many instances are extremely small. We further examine these cases and find that most of the components are concentrated on the types of \textit{CheckBox}, \textit{RadioButton}, \textit{Spinner}, and \textit{Switch}. For the component height distribution, Google Play apps present a concentration performance compared with F-Droid apps. 40dp is the most frequent height in commercial apps. The distribution range is relatively wide for F-Droid apps (i.e., concentrating between 30dp and 45dp). Also, similar to the width issues, several cases use 20dp height in F-Droid apps with serious touch target issues.

\smallskip
\noindent\fbox{
\parbox{0.95\linewidth}{
\noindent \textbf{Answer to RQ4.} We analyze the error patterns of the most frequent issues, and find (1) the low text and image contrast are caused by the wrong selection of color schema such as the foreground gray text on white background, and white image button above colorful background picture. (2) The small size of clickable components hinders users' usage and those issues are more serious in F-Droid apps than that of Google Play apps. But some \textit{touch target} issues are intentionally created for directing users to click the advertisements.
}}

\subsection{RQ5: Issue Fixing Analysis }\label{subsec:tracking}
{Due to the competitive market, the mobile development team frequently update their apps to gain the market share by releasing new features~\cite{chen2019storydroid}, fixing reported bugs~\cite{fan2018large, fan2018efficiently, liu2014characterizing, wei2016taming}, patching security bugs~\cite{chen2018mobile,chen2019ausera}, etc. However, using Alshayban et al.'s method cannot analyze the issue fixing status effectively and accurately due to the unsteady activity coverage of Monkey (flaky tests~\cite{thorve2018empirical, linares2017continuous, pecorelli2018testing, rubinov2018we}). Meanwhile, their fixing results are not manually validated, thus cannot conclude whether the previous detected issues are truly fixed. They found that 47\% of app updates improve the overall accessibility, 28\% of the updates impacted the overall accessibility negatively, and for the remaining 25\% overall accessibility levels remained the same~\cite{alshaybanaccessibility}. 

In this section, we aim to analyze the issue fixing status during app evolution by leveraging \ours. We randomly selected app package names crawled from Google Play, and collected the history versions of these apps from APKMonk~\cite{apkmonk} because Google Play only maintains the latest version. To minimize the side-effect caused by functionality addition and deletion when investigating the issue number changes during app evolution, we select the 3 latest versions of each app as the experimental subjects to observe whether the issues have been fixed from the aspect of accessibility improvement.
To this end, we collected 70 apps with 210 different versions, including some popular ones such as \textit{Booking}~\cite{booking} and \textit{Amazon Assistant}~\cite{amazon}. We do not investigate a large-scale dataset of apps because we need to manually cross-validate the issues on each page of each version.} Based on \ours, we collect the accessibility issue results for each version under the same experimental environment. After that, we manually compare the results among different versions for each app, including the number of issues detected in each version, the details about the issues, together with the reasons of issue number changing.

\begin{figure}
	\centering
	\includegraphics[width=0.425\textwidth]{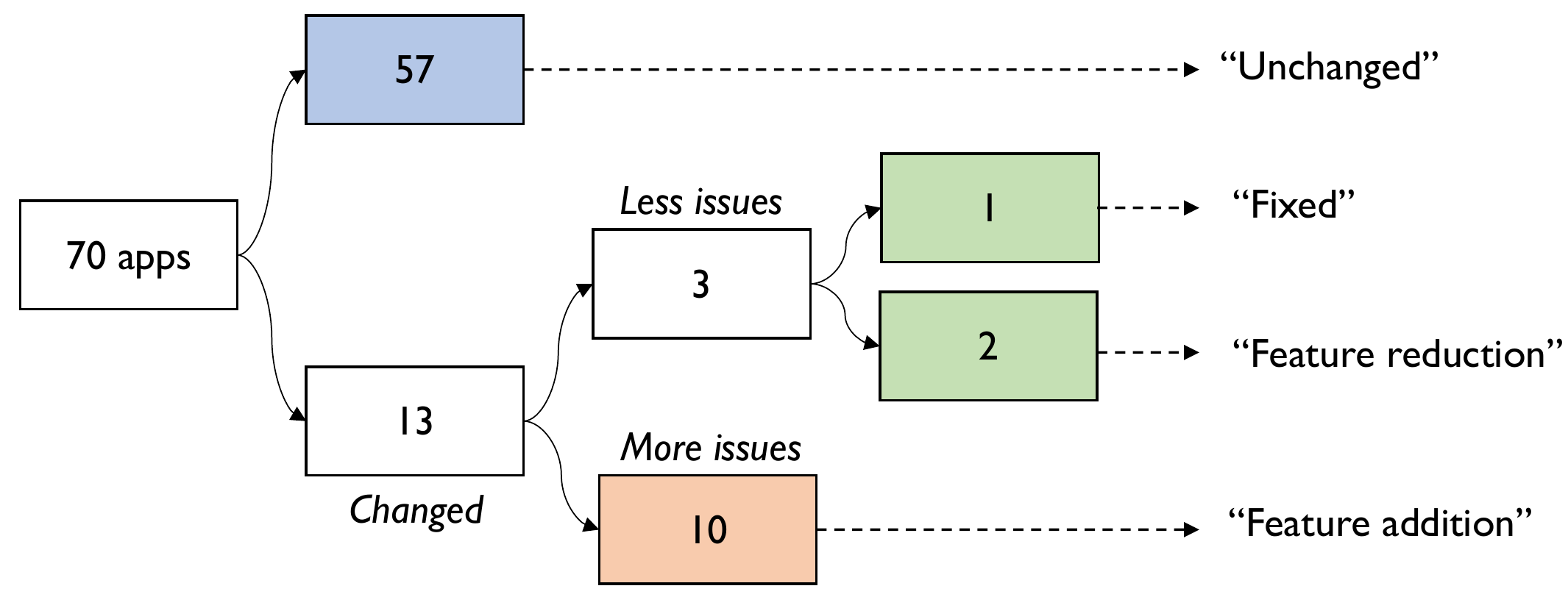}
	\caption{Issue fixing analysis in 70 apps with 210 versions. The number inside the box represents the number of apps.}
	\label{fig:evolution}
\end{figure} 

\begin{figure}
\centering
\subfigure[Example of accessibility issue number changes due to functionality deletion or addition]{%
	\includegraphics[width=0.4\textwidth]{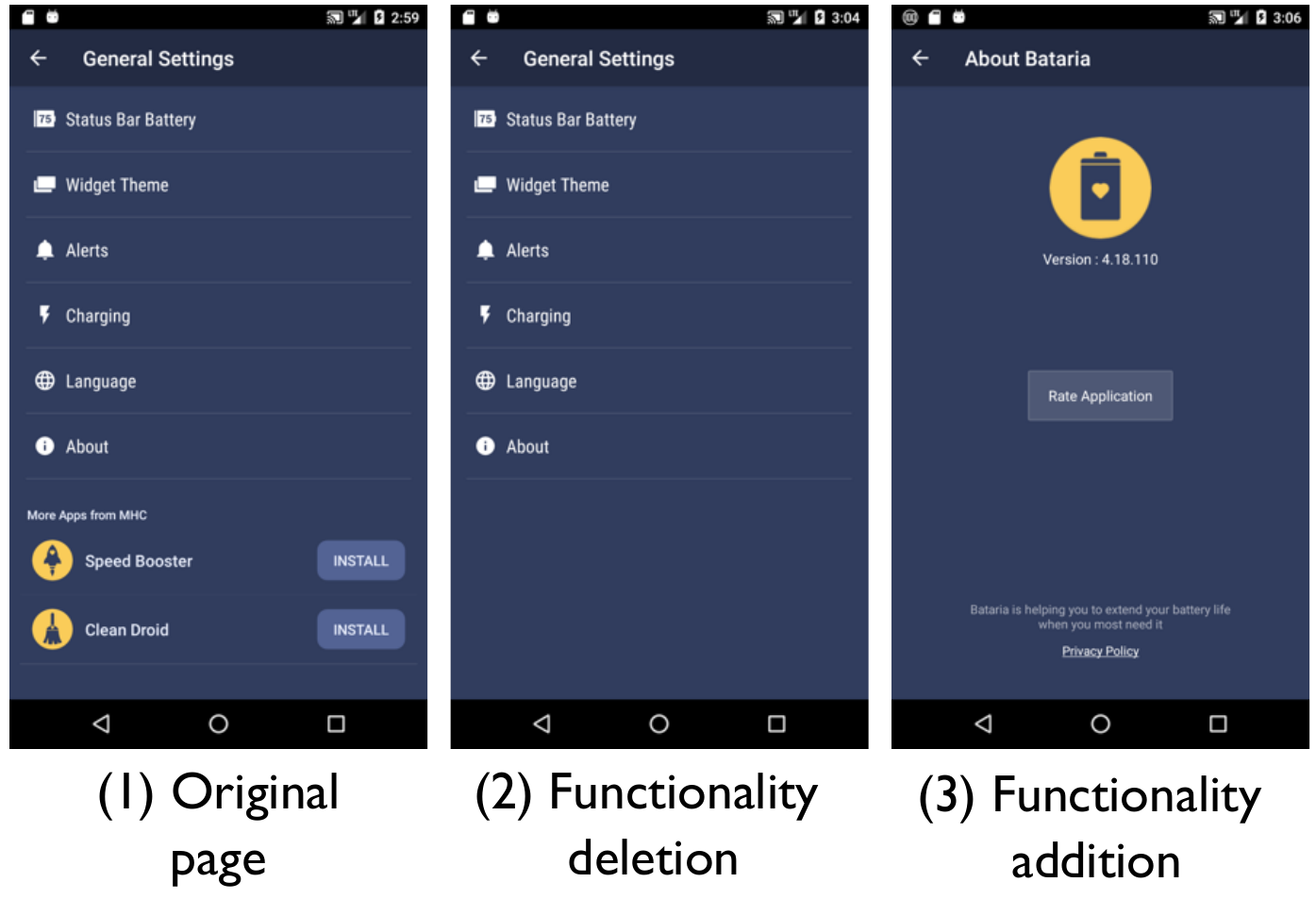}
	\label{fig:track_1}
}
\hfill
\subfigure[Example of accessibility issue fixed]{%
	\includegraphics[width=0.4\textwidth]{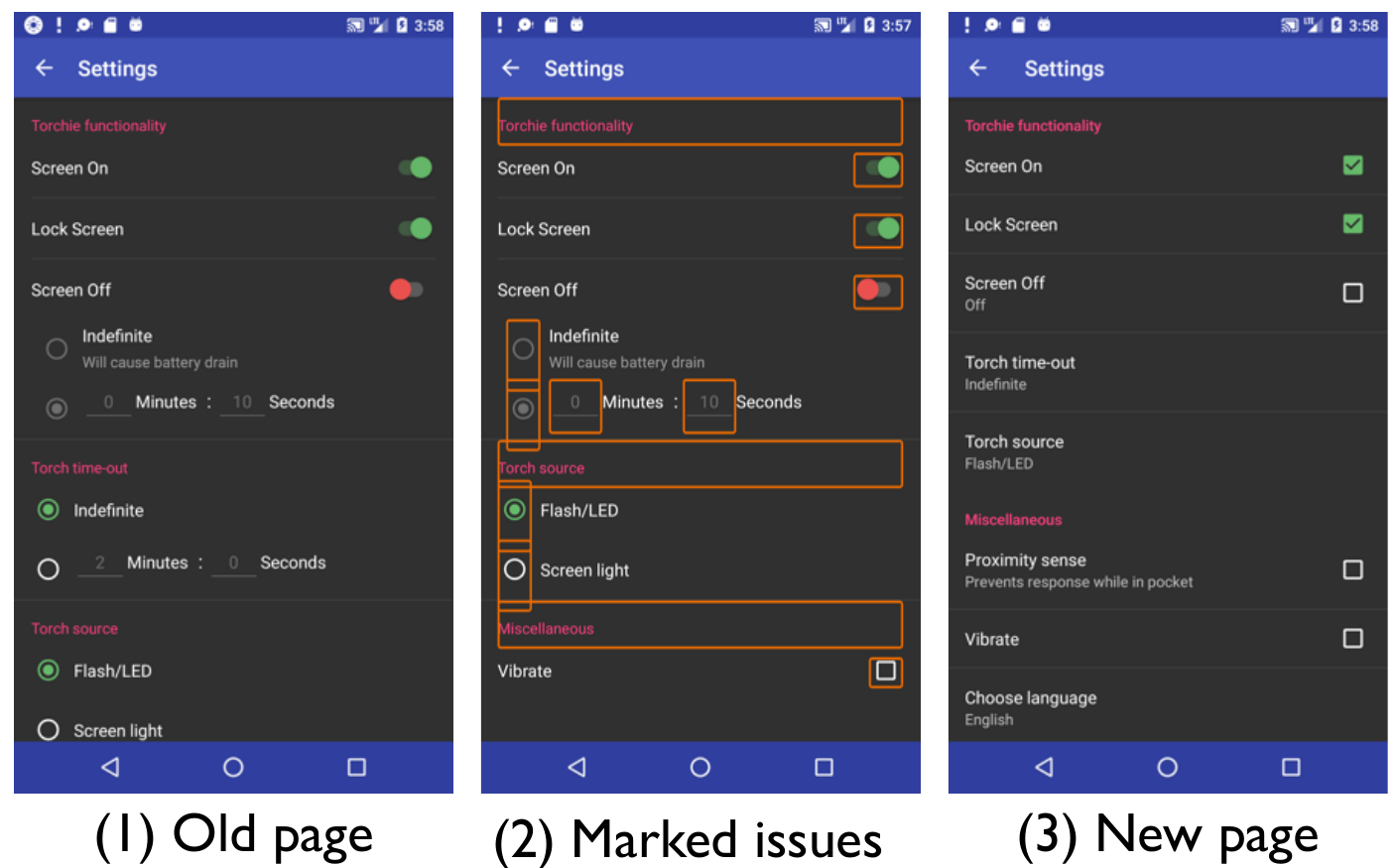}
	\label{fig:track_2}
}
\caption{Real examples of accessibility issue number changes}
\label{fig:resultEdit}
\end{figure}

Fig.~\ref{fig:evolution} shows the number of apps with different status. Among the 70 apps, we find that the number of issues across different versions is unchanged in 57 apps (81.43\%, marked blue in Fig.~\ref{fig:evolution}). The reasons for the ignorance is that either the development team do not locate these issue, or they are not motivated or knowledgeable enough to fix these issues. The number of issues changes in 13, and 10 (14.19\%, marked orange in Fig.~\ref{fig:evolution}) of them are detected with more issues during app updates. That is because of the new feature release accompanied with more screens, resulting in more issues. For example, an app description page (Fig.~\ref{fig:track_1} (3)) is added into this app, introducing 2 additional accessibility issues. Finally, there are only 3 apps (4.29\%) detected with less issues during their life-cycles. By observing their issue evolution, we find that the reason for the issue number decline in one app \textit{Battery Saver-Bataria Energy Saver} is that they delete some features (i.e., functionality module), hence two issues attached are removed. Fig.~\ref{fig:track_1} (1) shows two \textit{touch target} issues, and the corresponding ``fixing'' page deletes the functionality of \textit{``More Apps from MHC''}~\cite{battery} leading to the disappearance of the issues (Fig.~\ref{fig:track_1} (2)). The real issue fixing only occurs in an app named \textit{Torchie-Volume Button Torch}{~\cite{touch}}. In detail, one page in the old version (2016-05-18) contains 13 accessibility issues such as \textit{touch target}, \textit{item descriptions}, and \textit{text contrast} as seen in Fig.~\ref{fig:track_2} (1) and Fig.~\ref{fig:track_2} (2). By re-designing and re-implementing the UI in the new release version (i.e., version 2017-08-24), all of these issues are fixed by removing low-contrast text, adjusting the image color schema and adding a content description to the UI components in Fig.~\ref{fig:track_2} (3). 

To conduct a fair comparison, we also conduct experiments on the dataset used in \their~\cite{alshaybanaccessibility} for the multi-version experiment. We requested for the dataset from the authors and obtained 37 apps with 92 versions, based on which we run \ours to observe the issue fixing status and compare the results obtained from \their.
After manually analyzing the results, we find that most of the accessibility issues are remained in the multiple app versions to investigate the fixing status. The number of issues is unchanged in 21 apps (56.76\%). 10 of them (27.03\%) are detected with more issues due to adding new features along with version updates. Taking the app named \texttt{Word Cloud} (package name: ice.lenor.nicewordplacer.app) as an example, for its UI page (ice.lenor.nicewordplacer.app.MainActivity) of version 2.2.3, Xbot detects three more accessibility issues (i.e., Text contrast and Touch target) compared with the version 2.2.2. The reason is that the version 2.2.3 involves an advertisement on the top of screen. Another example is \texttt{Hairstyles step by step} (package name: com.piupiuapps.hairstyles), whose new version introduces more issues (i.e., Touch target issue) due to adding the text of ``Privacy Policy''. Only 6 apps (16.22\%) have less issues during version updates, where the developers delete some features instead of really fixing issues to improve the app accessibility. The overall result is consistent with the results on our dataset of 70 apps with 210 different versions.

\smallskip
\noindent\fbox{
\parbox{0.95\linewidth}{
\noindent \textbf{Answer to RQ5.} Analyzing the version history of selected apps indicates that the accessibility issues are rarely fixed by the development team. With the increase of app features, more issues are usually introduced. Some accessibility issues are fixed due to the reduction of features and only a few issues are intentionally fixed. Our results are different from the findings in~\cite{alshaybanaccessibility}, where they claimed apps become more accessible over time, with nearly half of app updates improving the overall accessibility, however without in-depth analysis on whether previous issues are truly fixed.}}

\section{Discussion}\label{sec:discussion}
The fine-grained and insightful findings demonstrate the great importance of issue collection for such an empirical study. These findings unveiled in Section~\ref{sec:new_study} may not be derived from the previous empirical studies due to the dataset with limited accessibility issues for each app. Last but not least, due to the low activity coverage of Monkey, issue fixing evolution cannot be accurately evaluated due to the flakiness nature of dynamic testing. Therefore, the 47\% fixing rate in~\cite{alshaybanaccessibility} might not be well validated. Such similar results would mislead the researchers, users, and developers in app accessibility. Finally, we, here, highlight that our study are from the perspective of accessibility issues themselves (i.e., issue level) and actually different and more in-depth compared with the previous studies at the screen level.

In the following, we first discuss implications of our study based on \ours and limitations of \ours, and motivates some future work.

\subsection{Design Implications}
\subsubsection{{For mobile app designers and developers}}
Despite having access to the accessibility guideline released by Android~\cite{android_access_guideline} and iOS~\cite{ios_access_guideline}, designers and developers may not understand them very well due to too abstract concepts and the lack of real examples. For example, it is not an easy task for designers to select color schema for not only highlighting the text, but also improving visual comfort, or increasing the size of the button. It is also difficult for developers to identify the views that a screen reader can focus and what descriptions should be added for supporting blind users. To help the development team better understand the accessibility issues, we are constructing a large-scale gallery~\cite{mysite} including both good GUI examples and ``negative'' GUIs with accessibility issues. Viewing these examples may help developers and designers who are not in the shoes of the disabled to learn both the good practice and also failure lessons about app accessibility. This gallery can complement with the accessibility guideline for elaborating the accessibility principles.

\subsubsection{{For mobile app release platform designers}}
Current mainstream app release platforms, such as Google Play~\cite{google_play}, support the app search by keywords and ratings, etc. However, as apps are more likely to be rated by users without disabilities, accessibility concerns from limited users tend to be diluted by other comments from users. 
Markets do not offer a mechanism to search apps based on their accessibility levels. Our tool can be used to assess the accessibility status of an app inferring an \textit{accessibility score} for it, similar to user ratings, which can be further used to rank the apps to facilitate people with disabilities to find more accessibility-friendly apps. Moreover, as our framework is capable of testing and evaluating the accessibility issues of a large number of apps efficiently, the app release platforms can leverage our framework to constantly evaluate the large volume of available apps and update the ranking of apps based on their accessibility as often as needed. Similar to previous Google's new mobile-friendly ranking algorithm that's designed to give a boost to mobile-friendly pages in Google's mobile search results~\cite{mobileSearch}, the app store can boost the accessibility-friendly apps in the app searching.

\subsection{Limitations and Future Work}
\textbf{First}, accessibility issues can happen even if all the GUI components are accessible. For example, a menu button may have good color contrast, the right size, and be positioned appropriately. However, the associated alternative text information can be inappropriate which can confuse a user with visual impairments \cite{Zhang:2017:IPR:3025453.3025846}. To detect such accessibility problems, the tool needs to be able to understand the appropriateness of the alternative text. Future work should examine how to integrate human judgments into the automated accessibility issue detection process. 
\textbf{Second}, our tool integrated the ability of Google Accessibility Test Framework~\cite{framework}, it detects accessibility issues based on a set of general accessibility rules, which are designed to cater for a set of common issues encountered by users with a wide range of disabilities. As a result, accessibility issues detected by our tool may be more than the issues that an individual user who only has a particular type of disability cares about. For example, a user with hearing impairments could care less about the accuracy of alternative texts, while a user with visual impairments would depend heavily on accurate alternative texts. Therefore, when using our tool to rate and rank the accessibility of mobile apps for users with disabilities, it is also important to consider the particular type of disability that users have and adapt the accessibility rating or ranking of mobile apps accordingly. Future work should examine more about how to dynamically customize mobile apps accessibility evaluation based on the particular types of disabilities that users have.
\textbf{Third}, our research, however, has not yet explored ways to recommend solutions to fix the detected accessibility issues or automatically fix these issues. Since this research has also created a large dataset of mobile apps with good and ``negative'' accessibility experience, future work could also examine ways to leverage the data, such as by training a deep learning model to provide app designers and developers with suggestions and examples to fix accessibility issues. 
Last, although the launched activity coverage (about 80\%) is much better than Monkey, it still does not achieve 100\%. The reasons are as follows. (1) Although we provide the Intent parameters, some activities still need to load other required data from local storage such as \textit{SQLite database} and remote server. Our tool cannot provide such types of data, which would cause errors. (2) Some apps require valid authentication, which means that they will check whether the app has been logged in successfully before launching pages.

\section{Conclusion} In this paper, we first highlight the challenges caused by the collected issue dataset in the previous empirical studies on app accessibility. We then propose an effective app exploration tool for automated accessibility testing of Android apps to mitigate the problem of issue data collection. Our tool achieves better performance when conducting accessibility testing. Based on our tool, we carry out a large-scale, in-depth investigation on 86,767 real accessibility issues and find that 88.99\% apps suffer from accessibility issues. We further unveil useful findings for app developers, designers, and research communities according to the results of the empirical study. Based on our findings, we further provide mobile app accessibility design implications for different stakeholders, such as app designers or developers, mobile app release platforms, and the mobile accessibility research community. Lastly, we highlight potential future research directions, including investigating methods to detect accessibility issues that still need human perception/intelligence to detect, to provide customized accessibility issues ratings based on users' specific disabilities, and to provide suggestions for fixing accessibility issues. Meanwhile, we released the dataset and the code of Xbot to facilitate the following works.

\section*{Acknowledgments}
This work was partially supported by the National Natural Science Foundation of China (Grant No. 62102284, 62102197).

\bibliographystyle{IEEEtran}
\bibliography{tse}
\begin{IEEEbiography}[{\includegraphics[width=1in,height=1.25in,clip,keepaspectratio]{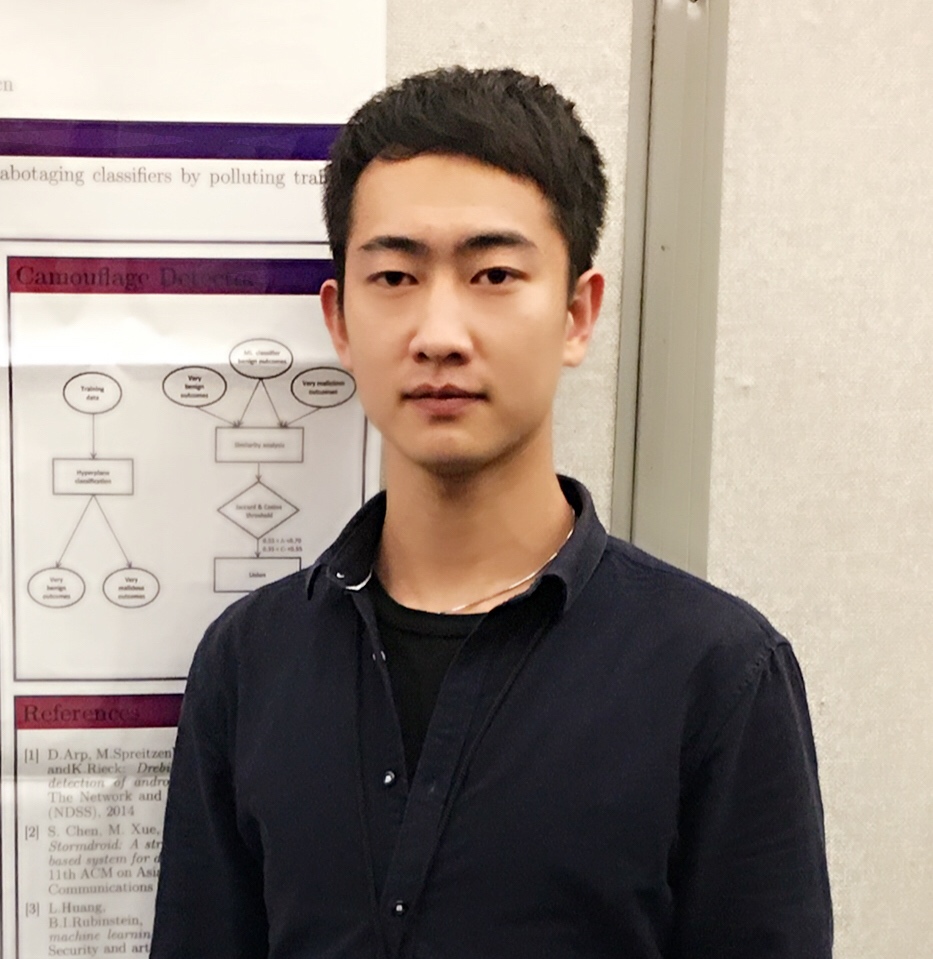}}]{Sen Chen} (Member, IEEE) is an Associate Professor in the College of Intelligence and Computing (School of Cybersecurity), Tianjin University, China. Before that, he was a Research Assistant Professor in the School of Computer Science and Engineering, Nanyang Technological University, Singapore. Previously, he was a Research Assistant of NTU from 2016 to 2019 and a Research Fellow from 2019-2020. He received his Ph.D. degree in Computer Science from School of Computer Science and Software Engineering, East China Normal University, China, in June 2019. His research focuses on Security and Software Engineering such as mobile security, AI security, open-source security, and intelligent development and testing. He has published broadly in top-tier security (IEEE S\&P, USENIX Security, CCS, IEEE TIFS, and IEEE TDSC) and software engineering venues including ICSE, FSE, ASE, ACM TOSEM, and IEEE TSE. More information is available on {\url{https://sen-chen.github.io/}.}
\end{IEEEbiography}

\begin{IEEEbiography}[{\includegraphics[width=1in,height=1.25in,clip,keepaspectratio]{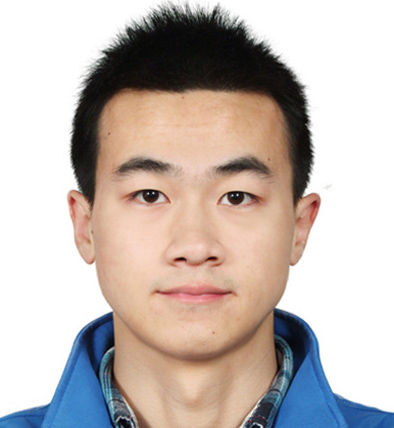}}]{Chen Chunyang} obtained his Ph.D. degree from School of Computer Science and Engineering, Nanyang Technological University (NTU), Singapore, and bachelor's degree from Beijing University of Posts and Telecommunications (BUPT), China, June 2014. He is a lecturer (a.k.a. Assistant Professor) in Faculty of Information Technology, Monash University, Australia. His research focuses on Mining Software Repositories, Text Mining, Deep Learning, and Human Computer Interaction.
\end{IEEEbiography}

\begin{IEEEbiography}[{\includegraphics[width=1in,height=1.25in,clip,keepaspectratio]{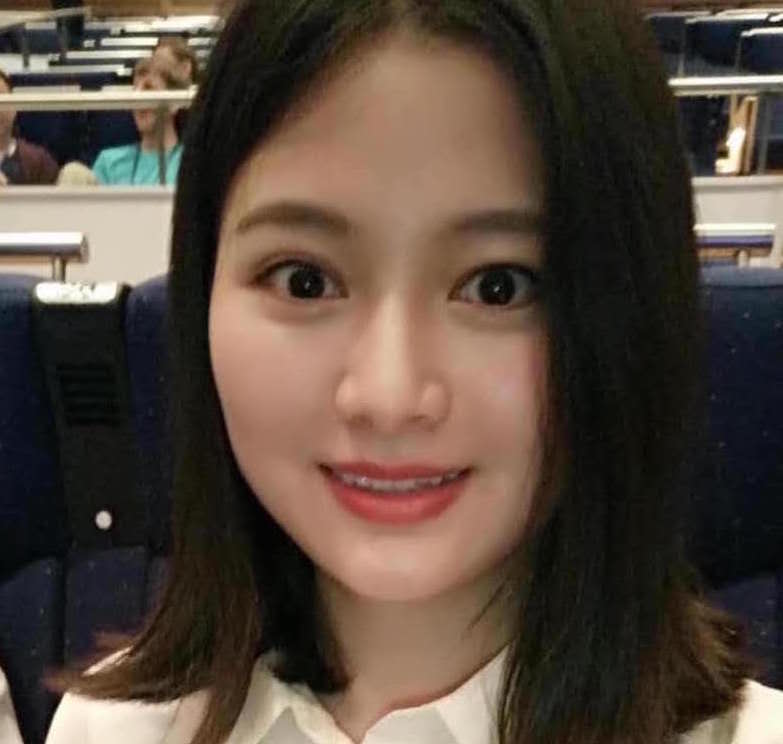}}]{Lingling Fan} is an Associate Professor in College of Cyber Science, Nankai University, China. She received her Ph.D and BEng degrees in computer science from East China Normal University, Shanghai, China in June 2019 and June 2014, respectively. In 2017, she joined Nanyang Technological University (NTU), Singapore as a Research Assistant and then had been as a Research Fellow of NTU since 2019. Her research focuses on program analysis and testing, software security, and Android and application analysis and testing. She got two ACM SIGSOFT Distinguished Paper Awards at ICSE 2018 and ICSE 2021. More information is available on~{\url{https://lingling-fan.github.io/}} 
\end{IEEEbiography}

\begin{IEEEbiography}[{\includegraphics[width=1in,height=1.25in,clip,keepaspectratio]{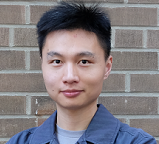}}]{Mingming Fan} is an Assistant Professor in the Computational Media and Arts Thrust and an Affiliated Assistant Professor in the Department of Computer Science and Engineering at The Hong Kong University of Science and Technology (HKUST) in Guangzhou and Clear Water Bay campuses respectively. He was an Assistant Professor at Rochester Institute of Technology from 2019 to 2021 and received a Ph.D. from the Department of Computer Science at the University of Toronto in 2019.

Dr. Fan leads the Accessible \& Pervasive User EXperience (APEX) Group to research in the field of Human-Computer Interaction and Accessibility. Specifically, his group applies user-centered design (UCD), AI, ML, VR/AR, visualization, sensing, and qualitative methods to 1) innovate User Experience (UX) Methodologies, 2) tackle Aging and Accessibility Challenges, and 3) Create Novel VR/AR Experience and Sensing Techniques. His research won Best Paper Award, Best Paper Honorable Mention Award, and Best Artifact Award from top-tier venues in HCI and Accessibility, such as ACM CHI, UbiComp, and ASSETS. 
More Info can be found at https://www.mingmingfan.com
\end{IEEEbiography}

\begin{IEEEbiography}[{\includegraphics[width=1in,height=1.25in,clip,keepaspectratio]{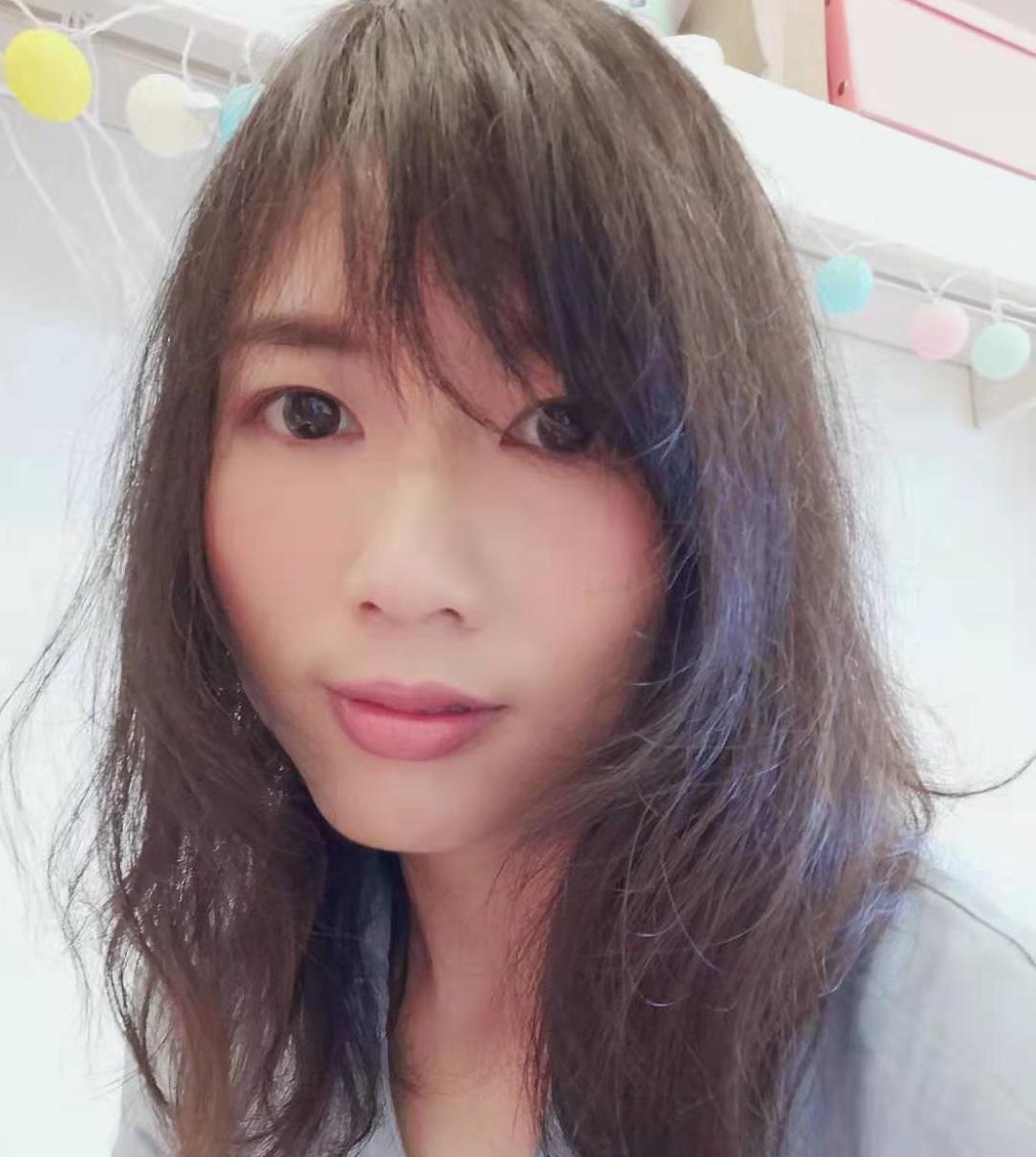}}]{Xian Zhan} received her BEng degree in Computer Science from Wuhan University, Hubei, China. Currently, she is a Ph.D candidate in the Department of Computing, the Hong Kong Polytechnic University. Her research interests include  program analysis, mobile privacy and security, NLP and machine learning. 
\end{IEEEbiography}

\vspace{-400pt}
\begin{IEEEbiography}[{\includegraphics[width=1in,height=1.25in,clip,keepaspectratio]{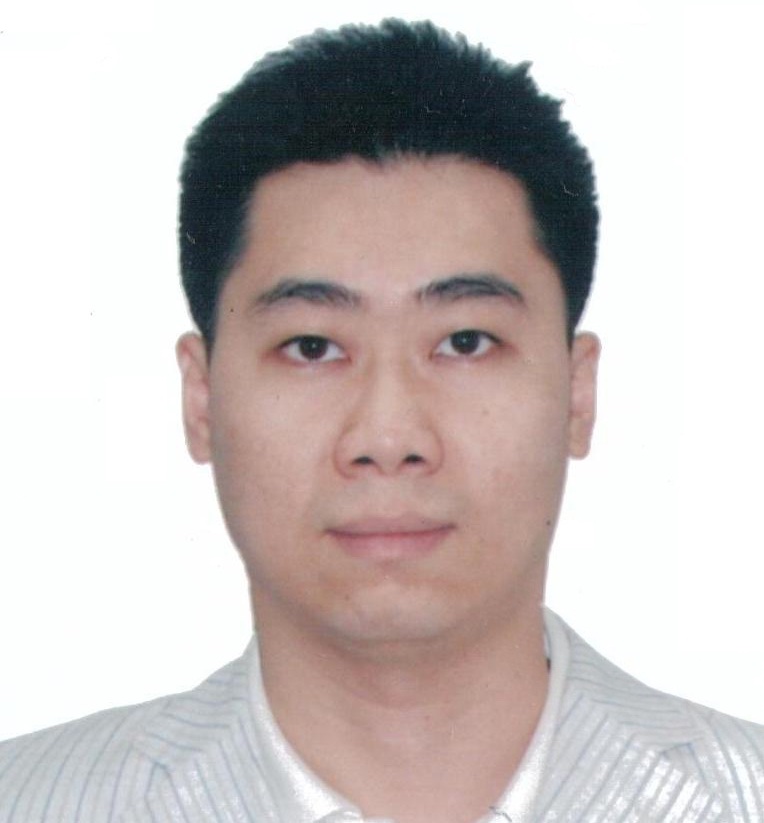}}]{Liu Yang}
graduated in 2005 with a Bachelor of Computing (Honours) in the National University of Singapore (NUS). In 2010, he obtained his PhD and started his post doctoral work in NUS, MIT and SUTD. In 2011, Dr. Liu is awarded the Temasek Research Fellowship at NUS to be the Principal Investigator in the area of Cyber Security. In 2012 fall, he joined Nanyang Technological University (NTU) as a Nanyang Assistant Professor. He is currently a full professor and the director of the cybersecurity lab in NTU.
              
He specializes in software verification, security and software engineering. His research has bridged the gap between the theory and practical usage of formal methods and program analysis to evaluate the design and implementation of software for high assurance and security. His work led to the development of a state-of-the-art model checker, Process Analysis Toolkit (PAT). By now, he has more than 300 publications and 6 best paper awards in top tier conferences and journals. With more than 20 million Singapore dollar funding support, he is leading a large research team working on the state-of-the-art software engineering and cybersecurity problems.
\end{IEEEbiography}
\end{document}